\PassOptionsToPackage{pdfpagelabels=false}{hyperref}
\documentclass[twoside,11pt,final]{entics} 
\usepackage{enticsmacro}
\usepackage{graphicx}
\usepackage[all]{xy}

\usepackage{amsmath}
\usepackage{array}
\usepackage{subcaption}
\usepackage{url}

\usepackage[table]{xcolor}
\usepackage{amsmath, amssymb, stmaryrd}
\usepackage{mathtools, url, multirow, makecell}

\usepackage{tikz}
\usepackage{tikz-cd}
\usepackage{makecell}
\usepackage{mathrsfs}
\usepackage{bookmark}

\makeatletter
\newcommand\newblock{\hskip .11em\@plus.33em\@minus.07em}
\makeatother

\newenvironment{mylemma}{\begin{lemma}}{\end{lemma}}
\newenvironment{myproposition}{\begin{proposition}}{\end{proposition}}
\newenvironment{mytheorem}{\begin{theorem}}{\end{theorem}}
\newenvironment{mycorollary}{\begin{corollary}}{\end{corollary}}
\newenvironment{mydefinition}{\begin{definition}}{\end{definition}}
\newenvironment{myremark}{\begin{remark}}{\end{remark}}
\newenvironment{myexample}{\begin{example}}{\end{example}}

\newcommand{\refdef}[1]{Def.~\ref{#1}}
\newcommand{\refprop}[1]{Prop.~\ref{#1}}
\newcommand{\reflem}[1]{Lem.~\ref{#1}}
\newcommand{\refthm}[1]{Thm.~\ref{#1}}
\newcommand{\refcor}[1]{Cor.~\ref{#1}}

\newcommand{\refexample}[1]{Example~\ref{#1}}

\newcommand{\refsec}[1]{Sec.~\ref{#1}}

\newcommand{\refitem}[1]{item~\ref{#1}}

\usetikzlibrary{cd}
\usetikzlibrary{math}

\newsavebox{\pullback}
\sbox\pullback{%
\begin{tikzpicture}%
\draw (0,0) -- (1ex,0ex);%
\draw (1ex,0ex) -- (1ex,1ex);%
\end{tikzpicture}}

\tikzset{heavy one/.style={blue!30}}
\tikzset{heavy/.style={color=blue!30, line width=5.2mm, rounded corners, line cap=round}}

\usetikzlibrary{positioning}
\usetikzlibrary{automata, arrows}

\usetikzlibrary{calc}
\usetikzlibrary{backgrounds,calc,trees,hobby}
\pgfdeclarelayer{background}
\pgfsetlayers{background,main}

\usepackage{algorithm}
\usepackage{algpseudocode}

\usepackage{wrapfig}

\newcommand{\sets}{\mathbf{Set}}

\newcommand{\idfunc}{\mathrm{Id}}
\newcommand{\idmorph}{\mathrm{id}}

\newcommand{\lift}[1]{\overline{#1}}
\newcommand{\pull}[1]{#1^{*}}
\newcommand{\push}[1]{{{#1}}_{*}}

\newcommand{\xmodal}{\mathsf{X}}

\newcommand{\umodal}{\mathsf{U}}

\newcommand{\emodal}{\mathsf{E}}
\newcommand{\amodal}{\mathsf{A}}
\newcommand{\wmodal}{\mathsf{W}}

\newcommand{\ap}{\mathrm{AP}}

\newcommand{\ttlog}{\mathsf{tt}}
\newcommand{\fflog}{\mathsf{ff}}

\newcommand{\Kleisli}[1]{\mathcal{K}{\kern-.2ex}\ell(#1)}

\newcommand{\sfmltopfml}[1]{{#1}}
\newcommand{\strength}{\mathsf{st}}

\newcommand{\nemppow}{\mathcal{P}^{+}}

\newcommand{\sem}[1]{\llbracket #1 \rrbracket} 

\newcommand{\unitInt}{{[0,1]}}
\newcommand{\seq}[2]{{#1}_{1},\dotsc,{#1}_{#2}}
\newcommand{\place}{\underline{\phantom{n}}\,} 

\newcommand{\pow}{\mathcal P}

\newcommand{\Bool}{\mathbf{2}}
\newcommand{\ttrue}{\mathrm{t{\kern-1.5pt}t}}
\newcommand{\ffalse}{\mathrm{f{\kern-1.5pt}f}}
\newcommand{\tuple}[1]{\langle#1\rangle}

\newcommand{\Kcomp}{\mathbin{\odot}}

\newcommand{\terminalobj}{\mathbf{1}}

\newcommand{\affinepart}[1]{{#1}_{\mathrm{a}}}

\newcommand{\mntnneighbor}{\mathcal{M}}

\newcommand{\citewithdata}[2]{\cite[#1]{#2}}

\newcommand{\Ord}{\mathbf{Ord}}

\newcommand{\StreamX}{X^{\omega}}
\newcommand{\traceoperator}{\mathsf{O}}

\newcommand{\umodalwith}[2]{{#1} \umodal {#2}}
\newcommand{\wmodalwith}[2]{{#1} \wmodal {#2}}

\newcommand{\LHS}{\text{LHS}}
\newcommand{\RHS}{\text{RHS}}

\newcommand{\PathFunctorX}{\Pi_{X}}

\newcommand{\PsiOperator}[2]{\Psi^{{#1}}_{{#2}}}
\newcommand{\PhiOperator}[2]{\Phi^{{#1}}_{{#2}}}
\newcommand{\PsiOperatorU}[1]{\Psi^{\umodal}_{{#1}}}
\newcommand{\PsiOperatorW}[1]{\Psi^{\wmodal}_{{#1}}}
\newcommand{\PhiOperatorW}[1]{\Phi^{\wmodal}_{{#1}}}
\newcommand{\PhiOperatorU}[1]{\Phi^{\umodal}_{{#1}}}
\newcommand{\PsiU}{\Psi^{\umodal}}
\newcommand{\PsiW}{\Psi^{\wmodal}}
\newcommand{\PhiW}{\Phi^{\wmodal}}
\newcommand{\PhiU}{\Phi^{\umodal}}

\newcommand{\traceoperatorT}[1]{\traceoperator_{{#1}}}

\newcommand{\ContMnd}{\mathcal{K}}
\newcommand{\contC}[1]{\Omega^{{#1}}}
\newcommand{\termType}[2]{{{#1}}^{{#2}}}

\newcommand{\HomObject}[2]{{{#1}}^{{{#2}}}}
\newcommand{\monotoneHom}[2]{[{{#1}}, {{#2}}]}

\newcommand{\monotoneContMndwith}[2]{\monotoneHom{\HomObject{{#1}}{{#2}}}{ {{#1}}}}
\newcommand{\monotoneContMnd}{\ContMnd^{\mathrm{m}}}

\newcommand{\affinemonotoneContMnd}{\ContMnd^{\mathrm{a}, \mathrm{m}}}

\newcommand{\strengthT}[1]{\strength^{{#1}}}

\newcommand{\canonModal}{\triangle}

\newcommand{\OsetMndOmega}[1]{\mathsf{T}^{#1}}
\newcommand{\OsetMnd}{\OsetMndOmega{\Omega}}

\newcommand{\affineOsetMndOmega}[1]{\affinepart{\OsetMndOmega{#1}}}
\newcommand{\affineOsetMnd}{\affineOsetMndOmega{\Omega}}

\newcommand{\llsem}{\llparenthesis}
\newcommand{\rrsem}{\rrparenthesis}
\newcommand{\Rsem}[1]{\llsem {#1} \rrsem}

\newcommand{\negfreeFixptML}{\mu\mathcal{L}}
\newcommand{\fullFixptML}{\mu\mathcal{L}^{\lnot}}
\newcommand{\TcoalgmodelFixptMLtuple}{(T,\Diamond,c,L)}
\newcommand{\BcoalgmodelFixptMLtuple}{(B,\Diamond,c,L)}
\newcommand{\coalgSemFixptMLsupsub}[3]{\sem{{#1}}^{{#2}}_{{#3}}}
\newcommand{\coalgSemFixptML}[1]{\coalgSemFixptMLsupsub{{#1}}{(B, \Diamond)}{c}}
\newcommand{\contmodelFixptMLtuple}{(c,L)}
\newcommand{\contSemFixptMLsupsub}[2]{\sem{{#1}}_{{#2}}}
\newcommand{\contSemFixptML}[1]{\contSemFixptMLsupsub{{#1}}{c}}

\newcommand{\opposite}{\mathsf{op}}

\newcommand{\polymorphicMonotoneCont}{\mathsf{K}^{\mathrm{m}}}

\newcommand{\exeOperator}{\traceoperator}
\newcommand{\exeOperatorofT}[1]{\exeOperator_{{#1}}}

\newcommand{\negfreeSfml}{\mathrm{s}\mathcal{L}_{\mathrm{CTL}^*}}
\newcommand{\negfreePfml}{\mathrm{p}\mathcal{L}_{\mathrm{CTL}^*}}
\newcommand{\negfreeCTL}{\mathcal{L}_{\mathrm{CTL}}}
\newcommand{\fullSfml}{\mathrm{s}\mathcal{L}^{\lnot}_{\mathrm{CTL}^*}}
\newcommand{\fullPfml}{\mathrm{p}\mathcal{L}^{\lnot}_{\mathrm{CTL}^*}}
\newcommand{\fullCTL}{\mathcal{L}^{\lnot}_{\mathrm{CTL}}}
\newcommand{\TcoalgmodelCTLtuple}{(T,\Diamond,c,L,u)}
\newcommand{\coalgSemCTLsupsub}[3]{\Rsem{{#1}}^{{#2}}_{{#3}}}
\newcommand{\coalgSemCTL}[1]{\coalgSemCTLsupsub{{#1}}{(T, \Diamond, u)}{c}}
\newcommand{\contmodelCTLtuple}{(c,L,u)}

\newcommand{\contSemCTLsupsub}[3]{\Rsem{{#1}}^{{#2}}_{{#3}}}
\newcommand{\contSemCTL}[1]{\contSemCTLsupsub{{#1}}{u}{c}}

\algnewcommand\algorithmicswitch{\textbf{switch}}
\algnewcommand\algorithmiccase{\textbf{case}}
\algnewcommand\algorithmicassert{\texttt{assert}}
\algnewcommand\Assert[1]{\State \algorithmicassert(#1)}%
\algdef{SE}[SWITCH]{Switch}{EndSwitch}[1]{\algorithmicswitch\ #1\ \algorithmicdo}{\algorithmicend\ \algorithmicswitch}%
\algdef{SE}[CASE]{Case}{EndCase}[1]{\algorithmiccase\ #1}{\algorithmicend\ \algorithmiccase}%
\algtext*{EndSwitch}%

\newcommand{\hobbyconvexpath}[2]{
[   
    create hobbyhullnodes/.code={
        \global\edef\namelist{#1}
        \foreach [count=\counter] \nodename in \namelist {
            \global\edef\numberofnodes{\counter}
            \node at (\nodename)
[draw=none,name=hobbyhullnode\counter] {};
        }
        \node at (hobbyhullnode\numberofnodes)
[name=hobbyhullnode0,draw=none] {};
        \pgfmathtruncatemacro\lastnumber{\numberofnodes+1}
        \node at (hobbyhullnode1)
[name=hobbyhullnode\lastnumber,draw=none] {};
    },
    create hobbyhullnodes
]
($(hobbyhullnode1)!#2!-90:(hobbyhullnode0)$)
\pgfextra{
  \gdef\hullpath{}
\foreach [
    evaluate=\currentnode as \previousnode using int(\currentnode-1),
    evaluate=\currentnode as \nextnode using int(\currentnode+1)
    ] \currentnode in {1,...,\numberofnodes} {
    \ifnum\currentnode=1\relax
    \xdef\hullpath{([closed=true]$(hobbyhullnode\currentnode)!#2!180:(hobbyhullnode\previousnode)$)
  ..($(hobbyhullnode\nextnode)!0.5!(hobbyhullnode\currentnode)$)}
    \else
    \xdef\hullpath{\hullpath
  ..($(hobbyhullnode\currentnode)!#2!180:(hobbyhullnode\previousnode)$)
  ..($(hobbyhullnode\nextnode)!0.5!(hobbyhullnode\currentnode)$)}
    \fi
    \ifx\currentnode\numberofnodes
    \else
    \xdef\hullpath{\hullpath
  ..($(hobbyhullnode\nextnode)!#2!-90:(hobbyhullnode\currentnode)$)}
    \fi
}
}
\hullpath
}

\def\conf{MFPS 2025} 	
\volume{5}			
\def\volu{5}			
\def\papno{13}			
\def\mydoi{16653}
\def\lastname{
    Kojima and C\^{i}rstea
    } 
\def\runtitle{Continuation Semantics} 
\def\copynames{R. Kojima, C. C\^{i}rstea}    
\begin{document}
\begin{frontmatter}
    \title{Continuation Semantics for Fixpoint Modal Logic and Computation Tree Logics} 						
  \author{Ryota Kojima\thanksref{a}\thanksref{myemail}}	
   \author{Corina C\^{i}rstea\thanksref{b}\thanksref{coemail}}		
   \address[a]{RIMS\\ Kyoto University\\				
    Kyoto, Japan}  							
   \thanks[myemail]{Email: \href{mailto:kojima@kurims.kyoto-u.ac.jp} {\texttt{\normalshape
   kojima@kurims.kyoto-u.ac.jp}}} 
  \address[b]{
    University of Southampton\\ Southampton, UK} 
  \thanks[coemail]{Email:  \href{mailto:cc2@ecs.soton.ac.uk} {\texttt{\normalshape
  cc2@ecs.soton.ac.uk}}}
\begin{abstract} 
    We introduce continuation semantics for both fixpoint modal logic (FML) and Computation Tree Logic* (CTL*), parameterised by a choice of branching type and quantitative predicate lifting. 
    Our main contribution is proving that they are equivalent to coalgebraic semantics, for all branching types. 
    Our continuation semantics is defined over coalgebras of the continuation monad whose answer type coincides with the domain of truth values of the formulas. 
    By identifying predicates and continuations, such a coalgebra has a canonical interpretation of the modality by evaluation of continuations. 
    We show that this continuation semantics is equivalent to the coalgebraic semantics for fixpoint modal logic. 
    We then reformulate the current construction for coalgebraic models of CTL*. 
    These models are usually required to have an infinitary trace/maximal execution map, characterized as the greatest fixpoint of a special operator. 
    Instead, we allow coalgebraic models of CTL* to employ non-maximal fixpoints, which we call execution maps. 
    Under this reformulation, we establish a general result on transferring execution maps via monad morphisms. 
    From this result, we obtain that continuation semantics is equivalent to the coalgebraic semantics for CTL*. 
    We also identify a sufficient condition under which CTL can be encoded into fixpoint modal logic under continuation semantics.
\end{abstract}
\end{frontmatter}

\section{Introduction}
\label{sec:introduction}
Modal and temporal logics are effective tools for specifying the behavior of various kinds of reactive systems, whose inner states are often invisible from the outside. 
Verifying specifications over such systems gives precious insight into how these systems behave, and what they produce in the next step and in the future.

In recent years, a coalgebraic approach to modelling and verifying reactive systems has shown a lot of promise, as it allows specification-description languages like \emph{coalgebraic modal logics}~\cite{Moss99,Pattinson03,Venema06} and \emph{coalgebraic temporal logics}~\cite{Cirstea14,Cirstea23,Cirstea11,Kojima+24} to be studied at a high level of generality, and verification techniques to be developed parametrically in the type of system and specification logic~\cite{CirsteaKupke23}.


The technical core of these coalgebraic modal and temporal logics is to interpret the modality using an ($\Omega$-valued) \emph{predicate lifting} of the branching-type endofunctor $B$. Such a predicate lifting is given by a monotone natural transformation of type $\contC{(\place)} \Rightarrow \contC{B(\place)}$ with $\Omega$ the domain of truth values for the formulas. This gives a categorical way to lift predicates over the state space to predicates over the set of ``$B$-computations'', which in turn allows modal formulas to be interpreted over $B$-coalgebras $c \colon X \rightarrow BX$. 

Our motivation is to represent these abstract constructs, including coalgebras, their branching types and the interpretation of the modality, in terms of the more concrete data structure of \emph{$\Omega$-valued continuations}~\cite{Reynolds93,Thielecke99}.
This intuition naturally comes if we transpose the type of the predicate lifting to the form $B(\place) \Rightarrow [\contC{(\place)} ,\Omega]$, which represents a natural transformation from the endofunctor $B$ to the monotone \emph{continuation monad}~\cite{Hyland+07,FuhrmannThielecke04} $\monotoneContMnd = [\contC{(\place)} ,\Omega]$ with answer type $\Omega$.
Via such a natural transformation $\iota$, every $B$-coalgebra $c$ is represented as the $\monotoneContMnd$-coalgebra $\iota \circ c$, whose successors are now just monotone maps, which consume $\Omega$-valued continuations over the state space and return their results in $\Omega$.

The \emph{continuation semantics} we propose in this paper interprets modal and temporal formulas over coalgebras of the continuation monad.
Since we can identify $\Omega$-valued predicates and $\Omega$-valued continuations, every $\monotoneContMnd = [\contC{(\place)} ,\Omega]$-coalgebra $c$ has a canonical interpretation of the modality given by mere \emph{evaluation} of $\Omega$-valued continuations with the function $c(x)$ at each state $x$.
By this ``built-in'' interpretation of the modality by evaluation, the coalgebras of the continuation monad do not only play the role of models of systems, which coalgebras are conventionally expected to play, but they also carry the information on predicate liftings at the same time.

$\monotoneContMnd$-coalgebras also come with a natural semantics for (path-based) temporal logics. Conventionally, coalgebraic models for temporal logics~\cite{Cirstea11,Kojima+24} require additional data to transform predicates over computation paths, representing temporal behaviors of system executions, to predicates over the set of states. When the branching type of systems is described by a monad $T$, this additional data is given by a Kleisli map $u \in \Kleisli{T}(X, \StreamX) = \sets(X, T\StreamX)$, linking states with their computation paths, usually obtained as the greatest fixpoint of a monotone operator $\exeOperatorofT{T}$ over the set $\Kleisli{T}(X, \StreamX)$ (where $X$ represents the state space).
For the monotone continuation monad $\monotoneContMnd$, producing such a fixpoint is an easy task, thanks to the rich ordered-structure on the set $\Kleisli{\monotoneContMnd}(X, \StreamX)$ inherited from the complete lattice $\Omega$.
Applying a general fixpoint theorem, we obtain not only the greatest fixpoint but also the least fixpoint for every $\monotoneContMnd$-coalgebra.
In this way, every $\monotoneContMnd$-coalgebra automatically becomes a model for temporal logic with the obtained map $u \in \Kleisli{\monotoneContMnd}(X, \StreamX)$ and moreover, the map $u$ provides a canonical interpretation of the path quantifier, by evaluation of $\Omega$-valued continuations with the function $u(x)$ at each state $x$ again.

After seeing several nice properties of $\monotoneContMnd$-coalgebras and continuation semantics over them, a natural question to ask is whether this semantics is equivalent to the original coalgebraic semantics for modal and temporal logic. The main technical contribution of this paper is to give a positive answer for this question.

We exemplify this equivalence with two well-known and highly expressive logics, namely \emph{fixpoint modal logic (FML)}~\cite{Kozen83} and (extended) \emph{Computation Tree Logic (CTL*)}~\cite{EmersonHalpern86}.
In the case of FML, we can incorporate every coalgebraic model as a $\monotoneContMnd$-coalgebra in the same way as aforementioned, via the natural transformation induced by the $\Omega$-valued predicate lifting. 
The highlight of the proof is to show that the transferred coalgebra $\iota \circ c$ has the same semantic information as the original model $c$.
This proof uses the key observation that every $\Omega$-valued predicate lifting for every endofunctor can be decomposed using the canonical interpretation of the modality, which, as we saw above, is obtained by evaluation.

As for the temporal logic CTL*, surprisingly, the current formulation of coalgebraic semantics is \emph{too restrictive} to be equivalent to continuation semantics.
This requires us to revisit the current definition of a coalgebraic model, so that the class of coalgebraic models becomes large enough to match up with the class of continuation models. 
Our main modification is that we allow \emph{every fixpoint} $u \in \Kleisli{T}(X, \StreamX)$ of the operator $\exeOperatorofT{T}$ to constitute a coalgebraic model for CTL*, and we call such a (possibly non-maximal) fixpoint a \emph{$T$-execution map}.
This removal of the maximality requirement might seem a subtle change, but 
the origin of this requirement goes back to the study of \emph{infinitary trace semantics}~\cite{Jacobs04a}, which to our knowledge was the first attempt to formalize ever-lasting behaviors of coalgebras.

As a result of this more general notion of execution map, 
we can show that a monad morphism transfers execution maps from its domain to its codomain.
When applied to our coalgebraic models for CTL*, 
the transferred execution map plays a key role for the equivalence of coalgebraic and continuation semantics for CTL*.


This paper introduces this novel, continuation-based approach to investigating behaviors of systems in a concrete and unified way.
While the main focus of this paper is to show the theoretical fundamentals of our approach,
we briefly sketch an application of our approach to analyzing the \emph{fixpoint characterization}~\cite{EmersonHalpern85} of CTL, which is essential for proving that CTL model checking can be done in linear time.

Our main contributions are summarized as follows.
\begin{itemize}
    \item 
    We introduce the continuation semantics~(\refdef{def:continuation-semantics-fixpoint-modal-logic}, \refdef{def:continuation-semantics-CTL}) for FML and CTL*, where in each case, the resulting model has its ``built-in'' interpretation of the modality $\Diamond$, respectively quantifier $\emodal$, characterized as evaluation of continuations~(\refprop{prop:evaluation-modality}, \refprop{prop:evaluation-quantifier}).
    \item 
    We show the equivalence of our continuation semantics and the coalgebraic semantics of these logics~(\refprop{prop:continuation-coalgebraic-sem-FixptML}, \refprop{prop:continuation-coalgebraic-sem-CTL}).
    In the case of CTL*, we obtain the equivalence from a general result relating execution maps for two different monads via monad morphisms~(\refprop{prop:over-monad-execution}).
    \item 
    We show that the minimal and maximal execution maps always exist for both $\monotoneContMnd$ and $\affinemonotoneContMnd$~(\refprop{prop:exe-operator-cont-monad}), and thus we can always extend every continuation model for FML to one for CTL* with one of these maps~(\refprop{prop:min-max-CTL-continuation-model}).
    \item We give a sufficient condition for the \emph{fixpoint characterization} and a weaker version of it to hold under continuation semantics~(\refthm{thm:fixpoint-characterization}).
\end{itemize}
This paper is organized as follows.
\refsec{sec:preliminaries} summarizes our conventions and preliminaries.
In \refsec{sec:Cont-sem-fixpt-modal-logic}, after recalling the coalgebraic semantics of FML, we introduce our continuation semantics and show the equivalence result.
In \refsec{sec:Cont-sem-CTL}, we reformulate the coalgebraic semantics for CTL*~\cite{Cirstea11,Kojima+24} with our novel notion of execution map, introduce our continuation semantics for CTL*, and prove the equivalence of the two semantics.
In \refsec{sec:min-max-unique-cont-execution}, we prove the existence of the minimal and maximal execution maps for the continuation monad.
In \refsec{sec:fixpt-characterization-Cont-sem-CTL}, we analyze the fixpoint characterization result under continuation semantics.

\section{Preliminaries}
\label{sec:preliminaries}
Throughout this paper, we follow the following conventions.
\begin{itemize}
    \item 
    Every functor treated in this paper is defined over the category $\sets$ of sets.
    We use the letters $B$ for an endofunctor, and $T$ or $S$ for monads. A \emph{$B$-coalgebra} for an endofunctor $B$ is a map of type $X \rightarrow BX$ for some set $X$. A $T$-coalgebra for a monad $T$ is also defined by seeing $T$ as a mere endofunctor.
    \item 
    We use the notations ${Y}^{X} := \sets(X, Y)$ and $[P, Q]$ for the set of monotone maps between ordered sets.
    A function $f \colon X \rightarrow Y$ is also represented using lambda notation $\lambda \termType{x}{X}. f(x)$, with the type superscript sometimes omitted. 
    We denote $\push{(\place)}$ and $\pull{(\place)}$, respectively, the post- and pre-compositions of function.
    \item We equip every monad $T$ with its \emph{monad strength}~\cite[Def.~5.2.9]{Jacobs16} given by the \emph{canonical monad strength}~\cite[Lem.~5.2.10]{Jacobs16}: for each $X,Y \in \sets$, the canonical monad strength $\strengthT{T} \colon X \times TY \rightarrow T(X \times Y)$ is defined by 
    $
    \strengthT{T} 
        := 
        \lambda\termType{(x,t)}{X \times TY}.\,
        T
        \bigl(
            \lambda\termType{y}{Y}.\, (x,y)
        \bigr)
        (t).
    $
    \item We fix a set $\ap$ of atomic propositions throughout this paper.
\end{itemize}

We consider the continuation monad whose answer type is given by a complete lattice (see \cite{cignoli+2000,DaveyPriestley02,gratzer11} for the further details of Lattice Theory). 
A \emph{complete lattice} is a poset $\Omega = (\Omega, \sqsubseteq)$ with every join (or equivalently, every meet).
Especially, it has the bottom and top elements $\bot = \bot_{\Omega}$ and $\top = \top_{\Omega}$. 
In this paper, we assume a complete lattice is always \emph{meet-} and \emph{join-continuous}: for every $I \in \sets$ and $a, b_i \in \Omega$ for $i \in I$, the equalities $a \sqcap \bigsqcup_{i \in I} b_{i}
        = \bigsqcup_{i \in I} a \sqcap b_{i}$ and $a \sqcup \bigsqcap_{i \in I} b_{i}
        = \bigsqcap_{i \in I} a \sqcup b_{i}$ hold.
A complete lattice $\Omega$ is called \emph{de Morgan} if it equips an involution map $\lnot$: explicitly, an involution map over $\Omega$ is a bijective monotone map $\lnot \colon \Omega \rightarrow \Omega^{\opposite}$ for the opposite lattice $\Omega^{\opposite} = (\Omega, \sqsubseteq^{\opposite})$ (we will identify $\lnot^{-1} = \lnot$).
The set of the booleans $\Bool$ is an obvious example of de Morgan (indeed Boolean) complete lattice.
Another example is the unit interval $\unitInt$: it is a complete lattice with $\sup$ as its join and $\inf$ as its meet. It also equips a de Morgan involution by the \emph{{\L}ukasiewicz negation} $\lnot r := 1-r$.

Hereafter, we fix a (possibly de Morgan) complete lattice $\Omega \in \sets$.

The \emph{plain continuation monad} $\ContMnd \colon \sets \rightarrow \sets$ with answer type $\Omega$ is defined by $\ContMnd (X) = \contC{\contC{X}}$ for $X \in \sets$ and $\ContMnd (f) = \lambda \termType{h}{\ContMnd X}.\, \lambda \termType{k}{\contC{Y}}.\, h (k \circ f) \colon \ContMnd X \rightarrow \ContMnd Y$ for $f \colon X \rightarrow Y$.
        Its monad unit and multiplication are given by $\eta_X = \lambda \termType{x}{X}.\, \lambda \termType{k}{\contC{X}}.\, k(x) \colon X \rightarrow \ContMnd X$ and $\mu_X = \lambda \termType{H}{\ContMnd \ContMnd X}.\, \lambda \termType{k}{\contC{X}}.\, H(\lambda \termType{h}{\ContMnd X}.\, h(k) ) \colon \ContMnd \ContMnd X \rightarrow \ContMnd X$, respectively.
We also define the monotone and affine sub-monads of the plain continuation monad: recall that a monad $T$ is \emph{affine} if its unit $\eta_{\terminalobj} \colon \terminalobj \rightarrow T\terminalobj$ is an isomorphism.
\begin{itemize}
    \item The \emph{monotone continuation monad} $\monotoneContMnd$ is defined by $\monotoneContMnd := \monotoneContMndwith{\Omega}{(\place)}$.
    \item The \emph{affine monotone continuation monad} $\affinemonotoneContMnd$ is defined by $\affinemonotoneContMnd(X) = \{ h \in \monotoneContMnd X \mid h(\lambda \place.\, a) = a \}$.\footnote{
        The affine monotone continuation monad $\affinemonotoneContMnd$ is indeed the \emph{affine part}~{\cite[Def.~4.5]{Jacobs94}} of $\monotoneContMnd$, which means $\affinemonotoneContMnd$ is the largest affine sub-monad of the monotone continuation monad.
    }
\end{itemize}
We introduce the ``polymorphic'' symbol $\polymorphicMonotoneCont$ to represent both of the two monads $\monotoneContMnd$ and $\affinemonotoneContMnd$ at once.
This symbol is used in the situations where we want to treat these monads in parallel.
We also
denote $\polymorphicMonotoneCont = \polymorphicMonotoneCont_{\Omega}$ when we emphasize the answer type $\Omega$ of the continuation monad $\polymorphicMonotoneCont$.
\begin{myexample}\label{eg:Boolean-continuations}
    When $\Omega = \Bool$, the monotone continuation monad $\monotoneContMnd_{\Bool}$ is called the \emph{monotone neighborhood (MN) monad}~\cite{HansenKupke04}, denoted $\mntnneighbor$.
    Under the identification of $\Bool^{X} = \pow X$, it is written as $\mntnneighbor X 
    = \{\mathcal{F} \in \pow\pow X \mid \text{$\mathcal{F}$ is upward closed} \}$.
    The MN monad $\mntnneighbor$ is used for coalgebraic formulations of the \emph{neighborhood semantics}~\cite{Montague70,Scott70} of modal logic since $\mntnneighbor$-coalgebras are exactly neighborhood frames.
\end{myexample}

We next recall the notion of \emph{monad morphism} between monads $S = (S, \eta^{S}, \mu^{S})$ and $T = (T, \eta^{T}, \mu^{T})$, which plays an important role throughout this paper:
a \emph{monad morphism} is a natural transformation $\iota \colon S \Rightarrow T$ which satisfies the equations
$\eta^T = \iota \circ \eta^S \colon \idfunc \Rightarrow T$ and
$\mu^T \circ \iota_T \circ S\iota = \mu^S \circ \iota \colon SS \Rightarrow T$.
We call a monad $S$ with a monad morphism $\iota \colon S \Rightarrow T$ a \emph{$T$-over-monad}.
        The monad $S$ is called a \emph{$T$-sub-monad} if the monad morphism $\iota$ is moreover injective.
Monad morphisms are compatible with the canonical monad strength and affineness of monads.
The following result comes from the definitions of monad morphism and affine-ness.
\begin{myproposition}\label{prop:monad-morphism-strength-affine}
    Let $\iota \colon S \Rightarrow T$ be a monad morphism.
    \begin{enumerate}
        \item For each $X, Y \in \sets$, the equation $\strengthT{T} \circ (\idmorph_X \times \iota_Y) = \iota_{X \times Y} \circ \strengthT{S}\colon X \times SY \rightarrow T(X \times Y)$ holds.
        \item When the monad $T$ is an affine monad, then so is the monad $S$.
    \end{enumerate}
\end{myproposition}
Trivially, the identity natural transformation $\idmorph \colon T \Rightarrow T$ is an invertible monad morphism for every monad $T$.
The inclusions between the above monads $\ContMnd, \monotoneContMnd,\affinemonotoneContMnd$ are also injective monad morphisms.

When $\Omega$ is a de Morgan complete lattice, there is another monad morphism between the affine monotone continuation monads $\affinemonotoneContMnd_{\Omega}$ and $\affinemonotoneContMnd_{\Omega^{\opposite}}$, induced from the involution $\lnot$.
\begin{myproposition}\label{prop:beta-monad-morphism}
    If $\Omega$ is a de Morgan complete lattice with its involution $\lnot$,
    the map 
    \begin{equation*}
            \beta_Y 
            := 
            \lambda h.\, 
            \lambda k.\, 
            \lnot h (\lnot \circ k) 
            \colon 
            \affinemonotoneContMnd_{\Omega^{\opposite}}Y \rightarrow \affinemonotoneContMnd_{\Omega}Y
    \end{equation*} 
    for each $Y \in \sets$ constitutes an invertible monad morphism $\beta \colon \affinemonotoneContMnd_{\Omega^{\opposite}} \Rightarrow \affinemonotoneContMnd_{\Omega}$.
\end{myproposition}
\begin{proof}
    We first see the map 
    $
    \beta_Y \colon \affinemonotoneContMnd_{\Omega^{\opposite}}Y \rightarrow \affinemonotoneContMnd_{\Omega}Y
    $
    is well-defined for each $Y \in \sets$.
    The only non-trivial part is to show $\beta_Y (h)$ is a monotone map for every $h \in \monotoneContMnd_{\Omega^{\opposite}}Y$: its affine-ness comes immediately.
    Let $k,k' \in \contC{Y}$ with $k \sqsubseteq k'$ and $h \in \monotoneContMnd_{\Omega^{\opposite}}Y$.
    It suffices to see $\lnot h(\lnot \circ k) \sqsubseteq \lnot h(\lnot \circ k')$ by the definition of the map $\beta_Y$.
    Since $k \sqsubseteq k'$ implies $k \sqsupseteq^{\opposite} k'$ and thus $\lnot \circ k \sqsubseteq^{\opposite} \lnot \circ k'$, we have $h(\lnot \circ k) \sqsubseteq^{\opposite} h(\lnot \circ k')$ by $h \in \monotoneContMnd_{\Omega^{\opposite}}Y$.
    Thus, we obtain $\lnot h(\lnot \circ k) \sqsupseteq^{\opposite} \lnot h(\lnot \circ k')$, which concludes $\lnot h(\lnot \circ k) \sqsubseteq \lnot h(\lnot \circ k')$.

    The map $\beta$ being monad morphism, namely, its naturality and compatibility with the monad unit and multiplication, comes from literal calculation.
\end{proof}

\section{Continuation Semantics for Fixpoint Modal Logic}\label{sec:Cont-sem-fixpt-modal-logic}
In this section, we introduce our continuation semantics for fixpoint modal logic (FML)~\cite{Kozen83}.
First, we recall the coalgebraic semantics of FML in the same manner as preceding work~\cite{Venema06,CirsteaKupkePattinson09}, using the categorical notion of \emph{predicate lifting}~\cite{Pattinson03} to interpret the modality.
We then note the well-known, seemingly folklore result on the bijective correspondence between predicate liftings and natural transformations into the continuation monad.
From this result, we obtain the canonical predicate lifting for the continuation monad $\polymorphicMonotoneCont$ induced from the identity map.
The continuation semantics introduced here uses this the canonical lifting to interpret the modality.
We show the equivalence of coalgebraic and continuation semantics for FML, using the result that this canonical lifting for $\polymorphicMonotoneCont$ decomposes every predicate lifting for \emph{every endofunctor}.

\subsection{Coalgebraic Semantics for Fixpoint Modal Logic}\label{subsec:syntax-fixpt-modal-logic}
Modal logic extended with least and greatest fixpoint operators $\mu$ and $\nu$, also called the \emph{modal $\mu$-calculus}, was first introduced by~\cite{Kozen83}.
We present here its full syntax and negation-free fragment.
\begin{mydefinition}\label{def:fixpoint-modal-logic}
    We define the set $\fullFixptML$, called the \emph{full fixpoint modal logic} (\emph{full FML}, for short), by the following grammar:
    \begin{equation*}
        \theta \in \fullFixptML ::= 
            u
            \mid
            p \in \ap
            \mid
            \lnot p
            \mid
            \ttlog
            \mid
            \fflog
            \mid
            \theta_1 \land \theta_2
            \mid
            \theta_1 \lor \theta_2
            \mid
            \Diamond \theta
            \mid
            \Box \theta  
            \mid
            \mu u.\, \theta(u)
            \mid
            \nu u.\, \theta(u)
    \end{equation*}
    where $u$ is the propositional variable, $\Diamond$ and $\Box$ are the mutually dual modalities, and $\mu$ and $\nu$ are the least and greatest fixpoint operators.
    
    We also define the set $\negfreeFixptML$, called the \emph{$\lnot$-free FML} as the set obtained by removing $\lnot p$ and $\Box$ from the above set $\fullFixptML$.
    We often call just \emph{FML formulas} for the formulas of both $\negfreeFixptML$ and $\fullFixptML$.
\end{mydefinition}

\begin{myremark}[negation as syntactic sugar]
    While negation is restricted to atomic propositions in the syntax of $\fullFixptML$ above, we can define the negation $\lnot \theta$ for every $\fullFixptML$ formula $\theta$ by inductively applying the de Morgan laws and the duality between the modalities $\Diamond, \Box$ and the fixpoint operators $\mu,\nu$: harmlessly, we can assume the syntactic equivalences $\lnot \Diamond \equiv \Box \lnot,\lnot \Box \equiv \Diamond \lnot$ and $\lnot \mu u.\, \theta(u) 
    \equiv 
    \nu u.\, \lnot \theta(\lnot u),\lnot \nu u.\, \theta(u) 
    \equiv 
    \mu u.\, \lnot \theta(\lnot u)$.
    Note that since $\theta(u)$ has only positive occurrences of the variable $u$, so has the formula $\lnot \theta(\lnot u)$.
    See \cite{GradelThomasWilke02} for further technical details.
\end{myremark}
The semantics for FML was originally defined over Kripke frames and later extended to neighborhood frames~\cite{EnqvistSeifanVenema15a}, each corresponding to Kripke semantics and neighborhood semantics for modal logic without fixpoints.
Both of these semantics are instances of \emph{coalgebraic semantics}~\cite{Venema06,CirsteaKupkePattinson09}, which is defined over $B$-coalgebras for an endofunctor $B$ with a \emph{predicate lifting}~\cite{Pattinson03} for $B$.\footnote{
    The original definition~\cite{Pattinson03} of predicate lifting, also employed by~\cite{Venema06,CirsteaKupkePattinson09}, was restricted to $\Bool$-valued case. Our definition here is slightly more general, in that we allow any complete lattice.
}
\begin{mydefinition}\label{def:Omega-predicate-lifting}
    Let $B$ be an endofunctor and $T$ be a monad.
    \begin{enumerate}
        \item 
        An \emph{$\Omega$-predicate lifting} for $B$
        is a natural transformation $\Diamond \colon \Omega^{(\place)}  \Rightarrow \Omega^{B(\place)}$ which is monotone with respect to the point-wise order on $\Omega^{Y}$ and $\Omega^{BY}$ for each $Y \in \sets$.
        \item 
        A \emph{cartesian $\Omega$-predicate lifting} for $T$
        is an $\Omega$-predicate lifting which is \emph{cartesian} as a natural transformation: a natural transformation $\Diamond \colon \Omega^{(\place)}  \Rightarrow \Omega^{T(\place)}$ is \emph{cartesian}~\citewithdata{Definition~3.1}{AguirreKatsumata20} if for each object $X$, 
        the equations $\pull{\eta} \circ \Diamond_X 
        = \idmorph_{\HomObject{\Omega}{X}} 
        \colon \HomObject{\Omega}{X} \rightarrow \HomObject{\Omega}{X}$ and
        $\pull{\mu} \circ \Diamond_X 
        = \Diamond_{TX} \circ \Diamond_X 
        \colon \HomObject{\Omega}{X} \rightarrow \HomObject{\Omega}{TTX}$
        hold.
    \end{enumerate}
\end{mydefinition}
\begin{mydefinition}\label{def:coalgebraic-model-fixpointML}
    An \emph{$\Omega$-valued coalgebraic model} is a tuple $\BcoalgmodelFixptMLtuple$ where $B$ is an endofunctor, $\Diamond \colon \Omega^{(\place)}  \Rightarrow \Omega^{B(\place)}$ is an $\Omega$-predicate lifting for $B$, $c \colon X \rightarrow BX$ is a $B$-coalgebra, and $L \colon \ap \rightarrow \HomObject{\Omega}{X}$ is a labeling function.
\end{mydefinition}
\begin{mydefinition}\label{def:coalg-semantics-fixpoint-modal-logic}
    Let $\BcoalgmodelFixptMLtuple$ be an $\Omega$-valued coalgebraic model. 
    For each $\negfreeFixptML$ formula $\theta$ with free variables $\seq{u}{m}$, 
    its interpretation $\coalgSemFixptML{\theta} \colon \bigl(\Omega^{X}\bigr)^{m}\rightarrow \Omega^{X}$ is defined by:
    \begin{align*}
        \coalgSemFixptML{u_i} (\vec{k})
        &:= k_i,
        &\qquad
        \coalgSemFixptML{p} (\vec{k})
        &:= L(p),
        \\
        \coalgSemFixptML{\ttlog} (\vec{k})
        &:= \lambda x.\, \top,
        &\qquad
        \coalgSemFixptML{\fflog} (\vec{k})
        &:= \lambda x.\, \bot,
        \\
        \coalgSemFixptML{\theta_1 \land \theta_2} (\vec{k})
        &:= 
        \coalgSemFixptML{\theta_1} (\vec{k})
        \sqcap
        \coalgSemFixptML{\theta_2} (\vec{k}),
        &\qquad
        \coalgSemFixptML{\theta_1 \lor \theta_2} (\vec{k})
        &:= 
        \coalgSemFixptML{\theta_1} (\vec{k})
        \sqcup
        \coalgSemFixptML{\theta_2} (\vec{k}),
        \\
        \coalgSemFixptML{\Diamond \theta} (\vec{k})
        &:= 
        \pull{c} \circ \Diamond_X
        \bigl(
            \coalgSemFixptML{\theta} (\vec{k})
        \bigr),
        \\
        \coalgSemFixptML{\mu u.\, \theta(u)} (\vec{k})
        &:= \mu k.\, \coalgSemFixptML{\theta(u)} (\vec{k}, k),
        &\qquad
        \coalgSemFixptML{\nu u.\, \theta(u)} (\vec{k})
        &:= \nu k.\, \coalgSemFixptML{\theta(u)} (\vec{k}, k)
    \end{align*}
    for $\vec{k} = (\seq{k}{m}) \in \bigl(\Omega^{X}\bigr)^{m}$.
    When $\Omega$ is a de Morgan complete lattice, the interpretation of $\fullFixptML$ formulas is also defined: in addition to the interpretation above, we define
    \begin{equation*}
        \coalgSemFixptML{\lnot p} (\vec{k})
        := \lnot \circ L(p),
        \qquad
        \coalgSemFixptML{\Box \theta} (\vec{k})
        := 
        \push{(\lnot)} \circ \pull{c} \circ 
        \Diamond_X
            \bigl(
                \lnot \circ \coalgSemFixptML{\theta} (\vec{k})
            \bigr).
    \end{equation*}
\end{mydefinition}

\subsection{Canonical Decomposition of Predicate Liftings}\label{subsec:predicate-lifting-Trinity}
As mentioned in Introduction~(\refsec{sec:introduction}), the following seemingly folklore result\footnote{
    The result appears in its partial forms in the literature in different contexts.
    For example, we can refer to \citewithdata{Proposition~1}{Hyland+07}, \citewithdata{Lemma~7}{JacobsMandemakerFurber16} and \citewithdata{Proposition~14}{Bonchi+23}.
}
is the starting point of our investigation for the continuation monad, and heavily used throughout this paper. 
\begin{myproposition}\label{prop:trinity-lemma}
    Let $B$ be an endofunctor and $T$ be a monad.
    \begin{enumerate}
        \item There is a bijective correspondence between the following:
        \begin{enumerate}
            \item a natural transformation $B \Rightarrow \monotoneContMnd$,
            \item an $\Omega$-predicate lifting $\Omega^{(\place)}  \Rightarrow \Omega^{B(\place)}$ for $B$.
        \end{enumerate}
        \item There is a bijective correspondence between the following:
        \begin{enumerate}
            \item a monad morphism $T \Rightarrow \monotoneContMnd$,
            \item a cartesian $\Omega$-predicate lifting $\Omega^{(\place)}  \Rightarrow \Omega^{T(\place)}$ of $T$.
        \end{enumerate}
    \end{enumerate}
\end{myproposition}
We only indicate concrete constructions of the first bijective correspondence: the full proof, including the cartesian case, comes immediately from this construction.
\begin{itemize}
    \item Given a natural transformation $\iota \colon B \Rightarrow \monotoneContMnd$, the map $\lambda \termType{k}{\HomObject{\Omega}{Y}}.\, \lambda \termType{t}{BY}.\, \iota_Y (t)(k) \colon \Omega^{Y}  \Rightarrow \Omega^{BY}$ for each $Y \in \sets$ gives an $\Omega$-predicate lifting for $B$.
    \item Given an $\Omega$-predicate lifting $\Diamond \colon \Omega^{(\place)}  \Rightarrow \Omega^{B(\place)}$ for $B$, the map $\lambda \termType{t}{BY}.\, \lambda \termType{k}{\HomObject{\Omega}{Y}}.\, \Diamond_Y (k)(t) \colon BY  \Rightarrow [\Omega^{Y}, \Omega] = \monotoneContMnd Y$ for each $Y \in \sets$ gives a natural transformation to $\monotoneContMnd$.
\end{itemize}
From \refprop{prop:monad-morphism-strength-affine}, the bijective correspondence for the cartesian case also applies to the affine monotone continuation monad $\affinemonotoneContMnd$.
\begin{myproposition}\label{prop:affine-trinity-lemma}
    Let $T$ be an affine monad.
    There is a bijective correspondence between the following:
        \begin{enumerate}
            \item a monad morphism $T \Rightarrow \affinemonotoneContMnd$,
            \item a cartesian $\Omega$-predicate lifting $\Omega^{(\place)}  \Rightarrow \Omega^{T(\place)}$ of $T$.
        \end{enumerate}
\end{myproposition}

The continuation monad $\polymorphicMonotoneCont$ has the cartesian predicate lifting induced from the identity.
\begin{mydefinition}\label{def:canonical-lifting}
    The \emph{canonical lifting}, denoted $\canonModal$, of $\polymorphicMonotoneCont = \polymorphicMonotoneCont_{\Omega}$ is the cartesian $\Omega$-predicate lifting induced from the identity natural transformation $\idmorph \colon \polymorphicMonotoneCont \Rightarrow \polymorphicMonotoneCont$ by~\refprop{prop:trinity-lemma}.
\end{mydefinition}
The canonical lifting $\canonModal$ of $\polymorphicMonotoneCont$ is the most basic one among all $\Omega$-predicate liftings of all endofunctors in that it decomposes every $\Omega$-predicate lifting.
Below, we formulate the statement for both the monotone and affine monotone cases in parallel using the polymorphic symbol $\polymorphicMonotoneCont$.
Throughout this paper, we will do similar parallel arguments as long as they are technically harmless.
\begin{myproposition}[canonical decomposition]\label{prop:canonical-decomposition-lifting}
    Let $B$ be an endofunctor, $\Diamond \colon \Omega^{(\place)}  \Rightarrow \Omega^{B(\place)}$ be an $\Omega$-predicate lifting for $B$ and $\iota \colon B \Rightarrow \polymorphicMonotoneCont$ be the natural transformation corresponding to $\Diamond$ by~\refprop{prop:trinity-lemma}, which means
    $\iota_Y = 
    \lambda \termType{t}{BY}.\,
    \lambda \termType{k}{\HomObject{\Omega}{Y}}.\, 
    \Diamond_Y (k)(t)$
    for each $Y \in \sets$.
    Then, the $\Omega$-predicate lifting $\Diamond$ for $B$ is decomposed as 
    \begin{equation*}
        \Diamond = \pull{\iota} \circ \canonModal
        \colon
        \Omega^{(\place)}  \Rightarrow \Omega^{\polymorphicMonotoneCont(\place)}
        \Rightarrow \Omega^{B(\place)}
    \end{equation*}
    via the canonical lifting $\canonModal \colon \Omega^{(\place)}  \Rightarrow \Omega^{\polymorphicMonotoneCont(\place)}$ of $\polymorphicMonotoneCont$.
    We call this decomposition of the $\Omega$-predicate lifting $\Diamond$ the \emph{canonical decomposition} of $\Diamond$.
\end{myproposition}
\begin{proof}
    The natural transformation $\iota$ is decomposed as $\iota = \idmorph \circ \iota \colon B \Rightarrow \polymorphicMonotoneCont \Rightarrow \polymorphicMonotoneCont$ via the identity natural transformation $\idmorph$ of $\polymorphicMonotoneCont$.
    This decomposed natural transformation corresponds to the $\Omega$-predicate lifting $\pull{\iota} \circ \canonModal$ by~\refprop{prop:trinity-lemma}.
    Since this correspondence is bijective and the corresponding lifting of the natural transformation $\iota$ is $\Diamond$ by definition, we have $\Diamond = \pull{\iota} \circ \canonModal$.
\end{proof}

\subsection{Continuation Semantics for Fixpoint Modal Logic}\label{subsec:Cont-sem-fixpt-modal-logic}
Our continuation semantics for FML is defined as an instantiation of the general coalgebraic semantics~(\refdef{def:coalg-semantics-fixpoint-modal-logic}) with the continuation monad $\polymorphicMonotoneCont$ and the canonical lifting $\canonModal$.
\begin{mydefinition}\label{def:continuation-semantics-fixpoint-modal-logic}
    \begin{enumerate}
        \item 
        An \emph{$\Omega$-valued continuation model} is a pair $\contmodelFixptMLtuple$ of
        a $\polymorphicMonotoneCont$-coalgebra $c \colon X \rightarrow \polymorphicMonotoneCont X$ and a labeling function $L \colon \ap \rightarrow \HomObject{\Omega}{X}$.
        \item 
        The \emph{continuation semantics} $\contSemFixptML{\place}$ of $\negfreeFixptML,\fullFixptML$ for a continuation model $\contmodelFixptMLtuple$ is defined as $\contSemFixptML{\place} := \coalgSemFixptMLsupsub{\place}{(\polymorphicMonotoneCont, \canonModal)}{c}$ for the $\Omega$-valued coalgebraic model $(\polymorphicMonotoneCont, \canonModal, c, L)$.
    \end{enumerate}
\end{mydefinition}
The key feature of the continuation semantics is that the interpretation of the modality $\Diamond$ is given by ``evaluation'', by the definition of the canonical lifting $\canonModal$.
\begin{myproposition}[modality is interpreted by evaluation]\label{prop:evaluation-modality}
    Let $\contmodelFixptMLtuple$ be a continuation model $\contmodelFixptMLtuple$. 
    For each FML formula $\theta$, we have
    \begin{equation*}
        \contSemFixptMLsupsub{\Diamond \theta}{c}(\vec{k}) 
        = 
        \lambda x.\, 
        c(x) 
        \bigl(
            \contSemFixptMLsupsub{\theta}{c}(\vec{k})
        \bigr).
    \end{equation*}
\end{myproposition}
The above presentation implies that each $\polymorphicMonotoneCont$-coalgebra itself carries both the data of a system and the lifting of predicates at the same time.
This suggests the special status of $\polymorphicMonotoneCont$-coalgebras among all coalgebras with various branching types in giving semantics to FML.
Indeed, the following result shows continuation semantics is strong enough to capture all semantic properties of FML.
\begin{myproposition}[continuation semantics is equivalent to coalgebraic semantics]\label{prop:continuation-coalgebraic-sem-FixptML}
    Let $\BcoalgmodelFixptMLtuple$ be an $\Omega$-valued coalgebraic model.
    \begin{enumerate}
        \item\label{item:continuation-coalgebraic-sem-FixptML-model} 
        The pair $(\iota_X \circ c, L)$ is a continuation model, where $\iota \colon B \Rightarrow \polymorphicMonotoneCont$ is the natural transformation induced from the $\Omega$-predicate lifting $\Diamond$ by~\refprop{prop:trinity-lemma}.
        \item\label{item:continuation-coalgebraic-sem-FixptML-semantics} 
        The interpretations $\contSemFixptMLsupsub{\theta}{\iota_X \circ c} = \coalgSemFixptML{\theta}$ coincide for every FML formula $\theta$.
    \end{enumerate}
\end{myproposition}
\begin{proof}
    \refitem{item:continuation-coalgebraic-sem-FixptML-model} is trivial.

    For \refitem{item:continuation-coalgebraic-sem-FixptML-semantics}, the proof goes inductively.
    The only non-trivial parts are the interpretations of $\Diamond \theta$ and $\Box \theta$.
    Assume $\contSemFixptMLsupsub{\theta}{\iota_X \circ c}(\vec{k}) = \coalgSemFixptML{\theta}(\vec{k})$ for $\vec{k} \in \bigl( \contC{X} \bigr)^m$ as the induction hypothesis.
    By the canonical decomposition $\Diamond = \pull{\iota} \circ \canonModal$ of~\refprop{prop:canonical-decomposition-lifting}
    and the definition of the coalgebraic semantics~(\refdef{def:coalg-semantics-fixpoint-modal-logic}), we have
    \begin{equation*}
        \coalgSemFixptML{\Diamond \theta}(\vec{k})
        =
        \pull{c} \circ \Diamond_X
        \bigl(
            \coalgSemFixptML{\theta} (\vec{k})
        \bigr)
        = \pull{c} \circ
        (\pull{\iota_X} \circ \canonModal_X)
        \bigl(
            \coalgSemFixptML{\theta} (\vec{k})
        \bigr)
        = \pull{(\iota_X \circ c)} \circ \canonModal_X
        \bigl(
            \contSemFixptMLsupsub{\theta}{\iota_X \circ c}(\vec{k})
        \bigr).
    \end{equation*}
    Since $\RHS$ of the above equation is equal to 
    $\coalgSemFixptMLsupsub{\Diamond \theta}{(\polymorphicMonotoneCont, \canonModal)}{\iota_X \circ c}(\vec{k}) = \contSemFixptMLsupsub{\Diamond \theta}{\iota_X \circ c}(\vec{k})$
    by the definitions of the coalgebraic and continuation semantics~(\refdef{def:coalg-semantics-fixpoint-modal-logic}, \refdef{def:continuation-semantics-fixpoint-modal-logic}), we conclude
    $\coalgSemFixptML{\Diamond \theta}(\vec{k}) = \contSemFixptMLsupsub{\Diamond \theta}{\iota_X \circ c}(\vec{k})$.
    We can show the $\Box \theta$ case by a similar calculation. 
\end{proof}
Recall that our continuation models are defined as a specific class of coalgebraic models (\refdef{def:continuation-semantics-fixpoint-modal-logic}).
The above result thus implies that every coalgebraic model is indeed a continuation model and vice versa.

\section{Continuation Semantics for Computation Tree Logics}
\label{sec:Cont-sem-CTL}
In this section, we formulate the coalgebraic semantics of Computation Tree Logic* (CTL*), mostly in line with preceding work~\cite{Cirstea11,Kojima+24} but with some modifications.
We first introduce the notion of \emph{$T$-execution map} of a $T$-coalgebra $c$ for a monad $T$, which we use to interpret the path quantifier of CTL*.
We then formulate our coalgebraic semantics and continuation semantics for CTL*, whose models have execution maps as new parameters.
We show the equivalence of these semantics using a general result on monad morphisms and execution maps.

\subsection{Execution Operator and Execution Maps}
\label{subsec:general-execution-maps}
Recall the \emph{Kleisli category} $\Kleisli{T}$ of a monad $T$ on $\sets$ (see \cite{Maclane71} for details).
Its objects are sets and its morphisms are defined as $\Kleisli{T}(Y,Y') := \sets(Y,TY')$ for $Y,Y' \in \sets$.
The composition $\Kcomp$ between morphisms is defined as 
    $
    f' \Kcomp f := \mu \circ Tf' \circ f \in \Kleisli{T}(Y,Y'')
    $
    for $f \in \Kleisli{T}(Y,Y')$ and $f' \in \Kleisli{T}(Y',Y'')$. 
    The identity morphism is defined as $\eta_Y \in \Kleisli{T}(Y,Y)$.
The category $\sets$ is embedded into the Kleisli category $\Kleisli{T}$ by the identity-on-objects functor $J \colon \sets \rightarrow \Kleisli{T}$ which maps a function $f \in \sets(Y,Y')$ to $Jf = \eta_{Y'} \circ f \in \sets(Y,TY')=\Kleisli{T}(Y,Y')$.

The set $\StreamX$ of paths over a set $X \in \sets$ can be characterized as the final coalgebra of the polynomial functor $\PathFunctorX := X \times \idfunc_{\sets}$, which we call the \emph{path functor}.
The final $\PathFunctorX$-coalgebra map $\zeta = \tuple{\zeta_1, \zeta_2} \colon \StreamX \xrightarrow{\cong} \PathFunctorX \StreamX = X \times \StreamX$ decomposes a path $\pi \in \StreamX$ into its head $\zeta_1(\pi) = \pi_0$ and tail $\zeta_1(\pi) = \pi^{+} := \pi_1 \pi_2 \dotsi$, meaning $\pi = \pi_0 \pi^{+}$.\footnote{
    Here $\pi_n$ represents the $n$-th element of a path $\pi$, and $x\pi \in \StreamX$ for an element $x \in X$ represents the concatenation of $x$ with $\pi$.
}
We denote $\lift{\PathFunctorX} \colon \Kleisli{T} \rightarrow \Kleisli{T}$ the Kleisli lifting of the path functor $\PathFunctorX$ for a monad $T$.
Explicitly, it is given by $\lift{\PathFunctorX}(Y) = \PathFunctorX(Y) = X \times Y$ for $Y \in \sets$ and $\lift{\PathFunctorX}(f) 
    = 
    \strength_{X, Y'} \circ (\idmorph_X \times f)$ for a Kleisli map $f\in \Kleisli{T}(Y,Y')$.


We define an \emph{execution map} of a $T$-coalgebra 
as a Kleisli map 
from its state space to the path space, which is additionally a fixpoint of a special operator, called \emph{execution operator}.
\begin{mydefinition}\label{def:maximal-trace-map}
    Let $T$ be a monad and $c \colon X \rightarrow T X$ be a $T$-coalgebra.
    \begin{enumerate}
        \item 
        The \emph{execution operator} $\exeOperatorofT{T,c} \colon \Kleisli{T}(X, \StreamX) \rightarrow \Kleisli{T}(X, \StreamX)$ of the $T$-coalgebra $c$ is a map defined by
        \begin{equation*}
            \exeOperatorofT{T,c} (u) 
            = J\zeta^{-1} 
            \Kcomp \lift{\PathFunctorX}(u) 
            \Kcomp 
            \bigl(
                \strengthT{T}_{X,X} \circ \tuple{\idmorph_X, c}
            \bigr)
        \end{equation*}
        for each $u \in \Kleisli{T}(X, \StreamX)$.
        We will omit its subscripts $T,c$ when they are clear from the context.
        \item 
        A \emph{$T$-execution map}, or just \emph{execution map}, of the $T$-coalgebra $c$ is a Kleisli map $u \in \Kleisli{T}(X, \StreamX)$ satisfying the equation $\exeOperatorofT{T,c} (u) = u$.
        \item 
        When the homset $\Kleisli{T}(X, \StreamX)$ is ordered, an execution map $u$ of the $T$-coalgebra $c$ is called the \emph{maximal execution map}~\cite{Cirstea11} if the map is the greatest one among all execution maps of $c$.
        We similarly define the \emph{minimal execution map}.
    \end{enumerate}
\end{mydefinition}
Intuitions behind the definition of the execution operator $\traceoperatorT{T,c}$ are as follows.
Let $x \in X$.
\begin{itemize}
    \item 
    The value 
    $\bigl(
        \strengthT{T}_{X,X} \circ \tuple{\idmorph_X, c}
    \bigr)(x)
    =
    \strengthT{T}_{X,X} \bigl( x, c(x) \bigr)
    \in T (X \times X)
    $ gives the successor $c(x)$ with the additional data of the current state $x$ on its left entry.
    \item 
    The map $\lift{\PathFunctorX}(u)$ applies the map $u$ only to the successors (the right entry), and does not touch the current state (the left entry).
    \item 
    The final part $J\zeta^{-1}$ concatenates the unchanged current state with the $u$-values of successors to produce an element in $T\StreamX$. (Recall that the isomorphism $\zeta$ decomposes a path to its head and tail.)
\end{itemize}
The equation $\exeOperatorofT{T,c} (u) = u$ implies directly applying the map $u$ to the current state $x$ is the same as applying the map to its all successors and then combining them as above.
\begin{myexample}\label{eg:pow-max-exe}
    When $T = \pow$, a $\pow$-coalgebra $c \colon X \rightarrow \pow X$ and a map $u \colon X \rightarrow \pow \StreamX$ induce
    \begin{equation*}
        \bigl(
        \strengthT{T}_{X,X} \circ \tuple{\idmorph_X, c}
        \bigr)(x) 
        = 
        \{ (x, x') \mid x' \in c(x) \},
        \quad
        \lift{\PathFunctorX}(u) (x, x')
        =
        \{ (x, \pi') \mid \pi' \in u(x') \},
        \quad
        J\zeta^{-1} (x, \pi')
        =
        \{ x\pi' \}.
    \end{equation*}
    The execution operator $\exeOperatorofT{\pow, c}$ can be calculated to
    $
    \exeOperatorofT{\pow,c} (u)(x) = 
        \{ x\pi' \in \StreamX \mid \exists x' \in c(x).\, \pi' \in u(x') \}
    $.
    The maximal $\pow$-execution map is well-known to be~\cite{Jacobs04a,UrabeHasuo18a} 
    \begin{equation*}
        x \mapsto u(x) = \{\pi \in X^{\omega} \mid \pi(0) = x \text{ and } \forall n \in \omega .\, \pi(n+1) \in c \bigl(\pi(n)\bigr) \}.
    \end{equation*}
    The minimal $\pow$-execution map is the trivial one: $\lambda\place.\, \emptyset$.
    When we replace the powerset monad $\pow$ with the non-empty powerset monad $\nemppow$, the maximal $\nemppow$-execution is known to be the same as above~\cite{Kojima+24}. 
\end{myexample}
Once defined abstractly, we now give a concrete formula to calculate the execution operator 
with only the defining data of a monad.
The following formula follows from the definition of the canonical strength map and applies to \emph{every} monad on the category $\sets$.
\begin{myproposition}\label{prop:concrete-formula-exe-operator}
    Let $T$ be a monad and $c \colon X \rightarrow T X$ be a $T$-coalgebra.
    The execution operator $\exeOperatorofT{T,c}$ can be calculated to
    $\exeOperatorofT{T,c} (u) 
    = 
    \lambda x.\, 
    T
    \bigl(
        \lambda \pi.\,
        x\pi
    \bigr)
    \bigl(
        (\mu_{\StreamX} \circ Tu \circ c) (x)
    \bigr)$
    for each $u \in \Kleisli{T}(X, \StreamX) = \sets(X, T\StreamX)$.
\end{myproposition}
\begin{myremark}
    The execution operator $\exeOperatorofT{T}$ defined here is just a specialization of the defining operator to obtain the more general coalgebraic notion of \emph{maximal trace} or \emph{infinitary trace}~\cite{Jacobs04a,UrabeHasuo18a}, which was introduced to capture ``ever-lasting'' behaviors of $TF$-coalgebras with a monad $T$ and an endofunctor $F$ under a \emph{distributive law} between them.
    A more restricted version of maximal trace, where the endofunctor $F$ and distributive law are fixed to a polynomial functor and the canonical distributive law, appear under the name of \emph{maximal execution map}~\cite{Cirstea11}.
    While our definition of execution map is in line with these preceding definitions, it differs from them in that we allow \emph{every fixpoint} of the execution operator.
    The modification is technically harmless in the formulation of coalgebraic semantics for CTL* of this paper (and also that of preceding work~\cite{Cirstea11,Kojima+24}).
    Moreover, our results are more clearly stated under this extended setting, as we will see in the later sections.
\end{myremark}

\subsection{Coalgebraic Semantics for Computation Tree Logics}
\label{subsec:coalg-sem-CTL}
Computation Tree Logic (CTL) and its two-sorted extension CTL* were introduced in~\cite{EmersonClarke82} and~\cite{EmersonHalpern86}, respectively.
We here present the $\lnot$-free and full versions of both of these logics.
\begin{mydefinition}\label{def:computation-tree-logic}
    We define the pair $(\fullSfml,\fullPfml)$ of sets, called the \emph{full extended Computation Tree Logic} (\emph{full CTL*}, for short),
    by the following (mutually inductive) grammar:
    \begin{align*}
        \psi \in \fullSfml ::=& 
            p \in \ap
            \mid
            \lnot p
            \mid
            \ttlog
            \mid
            \fflog
            \mid 
            \psi_1 \land \psi_2 
            \mid 
            \psi_1 \lor \psi_2 
            \mid \emodal \varphi
            \mid \amodal \varphi,
        \\
        \varphi \in \fullPfml ::=&
            \ttlog
            \mid
            \fflog
            \mid 
            \varphi_1 \land \varphi_2 
            \mid 
            \varphi_1 \lor \varphi_2
            \mid 
            \sfmltopfml{\psi}
            \mid 
            \xmodal \varphi 
            \mid 
            \umodalwith{\varphi_2}{\varphi_1}
            \mid 
            \wmodalwith{\varphi_1}{\varphi_2} 
    \end{align*}
    where $\emodal$ and $\amodal$ are the mutually dual path quantifiers.
    The \emph{$\lnot$-free CTL*} is defined as the pair $(\negfreeSfml,\negfreePfml)$ of sets obtained
    by removing $\lnot p$ 
    and $\amodal$ from the above set $\fullSfml$.
    We call the formulas of $\negfreeSfml/\fullSfml$ and $\negfreePfml/\fullPfml$, respectively, \emph{state formulas} and \emph{path formulas}.

    We also define the (strict) subset $\fullCTL \subsetneq \fullSfml$, called the \emph{full CTL} by the following grammar:
    \begin{align*}
        \psi \in \fullCTL ::=& 
            p \in \ap
            \mid
            \lnot p
            \mid
            \ttlog
            \mid
            \fflog
            \mid 
            \psi_1 \land \psi_2 
            \mid 
            \psi_1 \lor \psi_2 
            \mid
            \emodal\xmodal \psi 
            \mid 
            \emodal(\umodalwith{\psi_2}{\psi_1})
            \mid 
            \emodal(\wmodalwith{\psi_1}{\psi_2})
        \\
            &\mid
            \amodal\xmodal \psi 
            \mid 
            \amodal(\umodalwith{\psi_2}{\psi_1})
            \mid 
            \amodal(\wmodalwith{\psi_1}{\psi_2}).
    \end{align*}
    The \emph{$\lnot$-free CTL} $\negfreeCTL \subsetneq \negfreeSfml$ is also defined similarly to the $\lnot$-free CTL*.
    We just call \emph{CTL formulas} for both formulas of the $\lnot$-free and full CTL.
\end{mydefinition}
\begin{myremark}
    As a similar remark to the full FML, negation for the full CTL* and CTL can be extended to all formulas in an inductive manner, via the duality between $\emodal,\amodal$ and $\umodal, \wmodal$.
\end{myremark}
In our coalgebraic semantics for CTL*, models have the additional data of an execution map.
\begin{mydefinition}\label{def:coalgebraic-model-CTL}
    An \emph{$\Omega$-valued temporal coalgebraic model} is a tuple $\TcoalgmodelCTLtuple$ which consists of an $\Omega$-valued coalgebraic model~(\refdef{def:coalgebraic-model-fixpointML}) $\TcoalgmodelFixptMLtuple$ with a monad $T$ and a cartesian $\Omega$-predicate lifting $\Diamond$, and additionally a $T$-execution map $u \colon X \rightarrow T\StreamX$ of $c$.
\end{mydefinition}
\begin{mydefinition}\label{def:coalg-semantics-CTL}
    Let $\TcoalgmodelCTLtuple$ be an $\Omega$-valued temporal coalgebraic model. 
    For each $\negfreeSfml$ formula $\psi$ and $\negfreePfml$ formula $\varphi$,
   the interpretation $\coalgSemCTL{\psi} \in \Omega^{X}$ and $\coalgSemCTL{\varphi} \in \Omega^{\StreamX}$ are defined by:
    \begin{align*}
        \coalgSemCTL{p} 
        &:= L(p),
        &
        &
        \\
        \coalgSemCTL{\ttlog} 
        &:= \lambda x.\, \top,
        &\quad
        \coalgSemCTL{\fflog} 
        &:= \lambda x.\, \bot,
        \\
        \coalgSemCTL{\psi_1 \land \psi_2} 
        &:= 
        \coalgSemCTL{\psi_1} 
        \sqcap
        \coalgSemCTL{\psi_2},
        &\quad
        \coalgSemCTL{\psi_1 \lor \psi_2} 
        &:= 
        \coalgSemCTL{\psi_1} 
        \sqcup
        \coalgSemCTL{\psi_2},
        \\
        \coalgSemCTL{\emodal \varphi} 
        &:= 
        \pull{u} \circ \Diamond_{\StreamX}
        \bigl(
            \coalgSemCTL{\varphi} 
        \bigr)
        &
        &
    \end{align*}
    for state formulas and
    \begin{align*}
        \coalgSemCTL{\ttlog} 
        &:= \lambda \termType{\pi}{\StreamX}.\, \top,
        &\coalgSemCTL{\fflog} 
        &:= \lambda \pi.\, \bot,
        \\
        \coalgSemCTL{\varphi_1 \land \varphi_2} 
        &:= 
        \coalgSemCTL{\varphi_1} 
        \sqcap
        \coalgSemCTL{\varphi_2},
        &\coalgSemCTL{\varphi_1 \lor \varphi_2} 
        &:= 
        \coalgSemCTL{\varphi_1} 
        \sqcup
        \coalgSemCTL{\varphi_2},
        \\
        \coalgSemCTL{\sfmltopfml{\psi}}
        &:= \pull{\zeta_1} (\coalgSemCTL{\psi}),
        &\coalgSemCTL{\xmodal \varphi} 
        &:= 
        \pull{\zeta_2} 
        \bigl(
            \coalgSemCTL{\varphi} 
        \bigr)
        \\
        \coalgSemCTL{\umodalwith{\psi_2}{\psi_1}}
        &:= 
        \mu w.\, \coalgSemCTL{\psi_1} \sqcup 
        \Bigl(
            \coalgSemCTL{\psi_2} \sqcap 
            \bigl(
                \pull{\zeta_2}(w)
            \bigr)
        \Bigr),
        &
        &
        \\
        \coalgSemCTL{\wmodalwith{\psi_1}{\psi_2}}
        &:= 
        \nu w.\, \coalgSemCTL{\psi_1} \sqcap 
        \Bigl(
            \coalgSemCTL{\psi_2} \sqcup 
            \bigl(
                \pull{\zeta_2}(w)
            \bigr)
        \Bigr)
        &
        &
    \end{align*}
    for path formulas.
    When $\Omega$ is a de Morgan complete lattice, the interpretation of $\fullSfml$ formulas is also defined: in addition to the interpretation above, we define
    \begin{equation*}
        \coalgSemCTL{\lnot p} 
        := \lnot \circ L(p),
        \qquad
        \coalgSemCTL{\amodal \varphi}
        := 
        \push{(\lnot)} \circ \pull{u} \circ 
        \Diamond_{\StreamX}
            \bigl(
                \lnot \circ \coalgSemCTL{\varphi} 
            \bigr).
    \end{equation*}
    The interpretations of $\negfreeCTL$ and $\fullCTL$ are defined by restricting that of $\negfreeSfml$ and $\fullSfml$.
\end{mydefinition}
We can also explicitly write down the interpretation of the operators $\xmodal,\umodal,\wmodal$: we have
$\contSemCTL{\xmodal \sfmltopfml{\psi}} 
    = 
    \lambda \pi.\,
    \contSemCTL{\psi}(\pi_1)$,
$\contSemCTL{\umodalwith{\psi_2}{\psi_1}}
= 
\lambda \pi.\,
\bigsqcup_{n \in \omega}
\bigl(
        \bigsqcap_{m<n} 
        \contSemCTL{\psi_2}(\pi_m)
\bigr)
\sqcap
\contSemCTL{\psi_1}(\pi_n)$
and
$\contSemCTL{\wmodalwith{\psi_1}{\psi_2}}
= 
\lambda \pi.\,
\bigsqcap_{n \in \omega}
\bigl(
    \bigsqcup_{m<n} 
    \contSemCTL{\psi_1}(\pi_m)
\bigr)
\sqcup
\contSemCTL{\psi_2}(\pi_n)$
for state formulas $\psi, \psi_1, \psi_2$.\footnote{
    Note that we have $\zeta_1 \circ \zeta_2 (\pi) = (\pi^{+})_0 = \pi_1$ and that the meet and join operations are continuous over the complete lattice $\Omega$, since we assumed $\Omega$ to be always meet- and join-continuous.
}
This abstract semantics recovers the classical Kripke semantics of CTL* when we employ the maximal execution map presented in~\refexample{eg:pow-max-exe}.
See~\cite{Cirstea11,Kojima+24} for details.

\subsection{Continuation Semantics for Computation Tree Logics}
\label{subsec:Cont-sem-CTL}
The continuation semantics for CTL* is also a restriction of the coalgebraic one~(\refdef{def:coalg-semantics-CTL}) for this logic.
\begin{mydefinition}\label{def:continuation-semantics-CTL}
    \begin{enumerate}
        \item An \emph{$\Omega$-valued temporal continuation model} is a tuple $\contmodelCTLtuple$ where $\contmodelFixptMLtuple$ is an $\Omega$-valued continuation model and $u \colon X \rightarrow \polymorphicMonotoneCont \StreamX$ is an execution map of the $\polymorphicMonotoneCont$-coalgebra $c$.
        \item The \emph{continuation semantics} $\contSemCTL{\place}$ of CTL* (and CTL) for an $\Omega$-valued temporal continuation model $\contmodelCTLtuple$ is defined as $\contSemCTL{\place} := \coalgSemCTLsupsub{\place}{(\polymorphicMonotoneCont, \canonModal, u)}{c}$ for the $\Omega$-valued temporal coalgebraic model $(\polymorphicMonotoneCont, \canonModal, c, L, u)$.
    \end{enumerate}
\end{mydefinition}
The interpretation of the quantifier $\emodal$ is explicitly calculated by the definition of the canonical lifting $\canonModal$~(\refdef{def:canonical-lifting}), which is again given by \emph{evaluation} of continuations.
\begin{myproposition}[quantifier is interpreted by evaluation]\label{prop:evaluation-quantifier}
    Let $\contmodelCTLtuple$ be a temporal continuation model.
    For each path formula $\varphi$,
    we have 
    \begin{equation*}
        \contSemCTL{\emodal \varphi} 
        = 
        \lambda x.\,
        u(x)
        \bigl(
            \contSemCTL{\varphi}
        \bigr).
    \end{equation*}
\end{myproposition}
The following result, stating that monad morphisms transfer execution maps,  is crucial for proving the equivalence of the continuation and coalgebraic semantics for CTL*.
\begin{myproposition}[monad morphism transfers execution maps]\label{prop:over-monad-execution}
    Let $S,T$ be monads with a monad morphism $\iota \colon S \Rightarrow T$, and $c \colon X \rightarrow SX$ be an $S$-coalgebra.
    We assume that the homset $\Kleisli{T}(X,\StreamX)$ is a poset with an order $\sqsubseteq$ and the execution operator $\traceoperatorT{T, \iota_X \circ c}$ of $T$-coalgebra $\iota_X \circ c$ is a monotone map with respect to this order $\sqsubseteq$.
    Then, we have the following:
    \begin{enumerate}
        \item\label{item:over-monad-execution-order} 
        The set $\Kleisli{S}(X,\StreamX)$ equips the preorder $\sqsubseteq_{\iota}$ defined as: for each $u,v \in \Kleisli{S}(X,\StreamX)$, $u \sqsubseteq_{\iota} v$ holds if $\iota \circ u \sqsubseteq \iota \circ v$ holds.
        This preorder $\sqsubseteq_{\iota}$ is a partial order if $\iota$ is an injective monad morphism.
        \item\label{item:over-monad-execution-monotone} 
        The execution operator $\exeOperatorofT{T, \iota_X \circ c}$ restricts to $\exeOperatorofT{S, c}$ on the homset $\Kleisli{S}(X,\StreamX)$:
        \begin{equation}
            \label{diag:trace-operator-restriction}
            \begin{tikzcd}[ampersand replacement=\&, row sep=0.5cm]
                \Kleisli{T}(X,\StreamX)
                \ar[r, "\exeOperatorofT{T, \iota_X \circ c}"]
                \&
                \Kleisli{T}(X,\StreamX)
                \\
                \Kleisli{S}(X,\StreamX)
                \ar[r, "\exeOperatorofT{S, c}"]
                \ar[u,  "\push{(\iota_{\StreamX})}"]
                \&
                \Kleisli{S}(X,\StreamX)
                \ar[u,  "\push{(\iota_{\StreamX})}"]
            \end{tikzcd}
        \end{equation}
        and $\exeOperatorofT{S, c}$ is a monotone map with respect to the induced order $\sqsubseteq_{\iota}$.
        \item\label{item:over-monad-execution-fixpoint}
        A map $u \in \Kleisli{S}(X,\StreamX)$ is a pre/post-fixpoint of $\exeOperatorofT{S,c}$ if and only if $\iota_{\StreamX} \circ u \in \Kleisli{T}(X,\StreamX)$ is a pre/post-fixpoint of $\exeOperatorofT{T, \iota_{X} \circ c}$.
        Moreover, a map $u \in \Kleisli{S}(X,\StreamX)$ being a fixpoint of $\exeOperatorofT{S, c}$ implies $\iota_{\StreamX} \circ u \in \Kleisli{T}(X,\StreamX)$ being a fixpoint of $\exeOperatorofT{T, \iota_{X} \circ c}$.
        The converse also holds when $\iota$ is an injection.
        \item\label{item:over-monad-execution-inclusion}
        If $u$ is an $S$-execution map of $c$, the map $\iota_{\StreamX} \circ u$ is a $T$-execution map of $\iota_{X} \circ c$.
    \end{enumerate}
\end{myproposition}
\begin{proof}
    \refitem{item:over-monad-execution-order} is straightforward.

    \refitem{item:over-monad-execution-monotone} follows from the following diagram (we here erased obvious subscripts for readability): given $u \in \Kleisli{S}(X,\StreamX)$, we have
    \begin{equation*}
        \begin{tikzcd}[column sep=0.3cm]
            &
            X\times TX
            \ar[r, "\strengthT{T}"]
            &
            T(X \times X)
            \ar[rr, "T \bigl( \idmorph \times(\iota \circ u) \bigr)"]
            &
            &
            T(X \times T\StreamX)
            \ar[r, "T\strengthT{T}"]
            &
            TT(X \times \StreamX)
            \ar[r, "\mu^{T}"]
            &
            T(X \times \StreamX)
            \\
            &
            &
            &
            T (X \times S\StreamX)
            \ar[ru, "T(\idmorph \times \iota)"]
            &
            S(X \times T\StreamX)
            \ar[u, "\iota"]
            \ar[r, "S\strengthT{T}"]
            &
            ST(X \times \StreamX)
            \ar[u, "\iota_T"]
            &
            \\
            X
            \ar[r, "\tuple{\idmorph, c}"]
            \ar[ruu, "\tuple{\idmorph, \iota \circ c}"]
            &
            X\times SX
            \ar[r, "\strengthT{S}"]
            \ar[uu, "\tuple{\idmorph, \iota}"]
            &
            S(X \times X)
            \ar[rr, "S \bigl( \idmorph \times(\iota \circ u) \bigr)"]
            \ar[uu, "\iota"]
            &
            &
            S(X \times S\StreamX)
            \ar[r, "S\strengthT{S}"]
            \ar[lu, "\iota"]
            \ar[u, "S(\idmorph \times \iota)"]
            &
            SS(X \times \StreamX)
            \ar[r, "\mu^{S}"]
            \ar[u, "S\iota"]
            &
            S(X \times \StreamX).
            \ar[uu, "\iota"]
        \end{tikzcd}
    \end{equation*} 
    Note that the top and bottom path of the above diagram are $\exeOperatorofT{T} \circ \push{\iota} (u)$ and $\push{\iota} \circ \exeOperatorofT{S}(u)$, respectively.
    Most squares in the diagram follows immediately from naturality of $\iota$ and $\strengthT{S},\strengthT{T}$.
    Recall that the strength of the $T$-over-monad $S$ is inherited from that of $T$~\refprop{prop:monad-morphism-strength-affine} (i.e.\ $\strengthT{T} \circ (\idmorph \times \iota)= \iota \circ \strengthT{S}$).
    The right-most square comes from the definition of monad morphism.

    For \refitem{item:over-monad-execution-fixpoint}, if $u$ is a post-fixpoint of $\exeOperatorofT{S}$, $\iota \circ u$ is immediately a post-fixpoint of $\exeOperatorofT{T}$ from \refitem{item:over-monad-execution-monotone}.
    Conversely,
    if $\iota \circ u$ is a post-fixpoint of $\exeOperatorofT{T}$, We have $\iota \circ u
    \sqsubseteq \exeOperatorofT{T}(\iota \circ u)
    = \iota \circ \exeOperatorofT{S}(u)$ by~diagram~\ref{diag:trace-operator-restriction} in~\refitem{item:over-monad-execution-monotone}.
    Thus, we conclude $u
    \sqsubseteq_{\iota} \exeOperatorofT{S}(u)$ by the definition of the induced order $\sqsubseteq_{\iota}$ of~\refitem{item:over-monad-execution-order}.
    The pre-fixpoint case is treated similarly.
    For the fixpoint case, since a fixpoint is both post- and pre-fixpoint, the above argument implies that $\iota \circ u$ is a fixpoint of $\exeOperatorofT{T, \iota \circ c}$.
    The converse is also true if the induced order $\sqsubseteq_{\iota}$ is a partial order.

    Finally, \refitem{item:over-monad-execution-inclusion} immediately follows from \refitem{item:over-monad-execution-fixpoint}.
\end{proof}
When applied to $\polymorphicMonotoneCont$-over-monads, this result enables us to transform every temporal coalgebraic model to a temporal continuation model.
\begin{myproposition}[continuation semantics is equivalent to coalgebraic semantics]\label{prop:continuation-coalgebraic-sem-CTL}
    Let $\TcoalgmodelCTLtuple$ be an $\Omega$-valued temporal coalgebraic model.
    \begin{enumerate}
        \item\label{item:continuation-coalgebraic-sem-CTL-model} 
        The tuple $(\iota_X \circ c, L, \iota_{\StreamX} \circ u)$ is an $\Omega$-valued temporal continuation model, where $\iota \colon T \Rightarrow \polymorphicMonotoneCont$ is the monad morphism corresponding to the cartesian $\Omega$-valued predicate lifting $\Diamond$.
        \item\label{item:continuation-coalgebraic-sem-CTL-semantics} 
        The interpretations $\contSemCTLsupsub{\place}{\iota_{\StreamX} \circ u}{\iota_X \circ c} = \coalgSemCTL{\place}$ coincide for every CTL* formula.
    \end{enumerate}
\end{myproposition}
\begin{proof}
    For \refitem{item:continuation-coalgebraic-sem-CTL-model}, since $\iota$ is a monad morphism, $\iota_{\StreamX} \circ u$ is an execution map of the $\polymorphicMonotoneCont$-coalgebra $\iota_X \circ c$ by \refitem{item:over-monad-execution-inclusion} of \refprop{prop:over-monad-execution}.

    For \refitem{item:continuation-coalgebraic-sem-CTL-semantics}, the proof goes inductively.
    The only non-trivial part is the interpretation of $\emodal \varphi$:
    the $\amodal \varphi$ case goes similarly.
    Assume $\contSemCTLsupsub{\varphi}{\iota_{\StreamX} \circ u}{\iota_X \circ c} = \coalgSemCTL{\varphi}$ for a path formula $\varphi$ as the induction hypothesis.
    The decomposition $\Diamond = \pull{\iota} \circ \canonModal$ of~\refprop{prop:canonical-decomposition-lifting}
    and the definition of the coalgebraic semantics~(\refdef{def:coalg-semantics-CTL}) induce
    \begin{equation*}
        \coalgSemCTL{\emodal \varphi}
        =
        \pull{u} \circ \Diamond_{\StreamX}
        \bigl(
            \coalgSemCTL{\varphi}
        \bigr)
        = \pull{u} \circ
        (\pull{\iota_{\StreamX}} \circ \canonModal_{\StreamX})
        \bigl(
            \coalgSemCTL{\varphi}
        \bigr)
        = \pull{(\iota_{\StreamX} \circ u)} \circ \canonModal_{\StreamX}
        \bigl(
            \contSemCTLsupsub{\varphi}{\iota_{\StreamX} \circ u}{\iota_X \circ c}
        \bigr).
    \end{equation*}
    Since $\RHS = \coalgSemCTLsupsub{\emodal \varphi}{(\polymorphicMonotoneCont, \canonModal, \iota_{\StreamX} \circ u)}{\iota_X \circ c} = \contSemCTLsupsub{\emodal \varphi}{\iota_{\StreamX} \circ u}{\iota_X \circ c}$, we conclude $\coalgSemCTL{\emodal \varphi} = \contSemCTLsupsub{\emodal \varphi}{\iota_{\StreamX} \circ u}{\iota_X \circ c}$.
\end{proof}

\section{Minimal and Maximal Execution Maps for the Continuation Monad $\polymorphicMonotoneCont$}\label{sec:min-max-unique-cont-execution}
In this section, we further investigate the $\polymorphicMonotoneCont$-execution maps and continuation models.

The first result is the existence of minimal and maximal $\polymorphicMonotoneCont$-execution maps.
\begin{myproposition}[minimal and maximal $\polymorphicMonotoneCont$-execution map]\label{prop:exe-operator-cont-monad}
    \begin{enumerate}
        \item\label{item:exe-operator-cont-monad-1} 
            Let $c \colon X \rightarrow \polymorphicMonotoneCont X$ be a $\polymorphicMonotoneCont$-coalgebra.
            The execution operator $\exeOperatorofT{\polymorphicMonotoneCont, c}\colon \Kleisli{\polymorphicMonotoneCont}(X,\StreamX) \rightarrow \Kleisli{\polymorphicMonotoneCont}(X,\StreamX)$ for the continuation monad $\polymorphicMonotoneCont$ is given by
            \begin{equation*}
                \exeOperatorofT{\polymorphicMonotoneCont, c}(u)(x) 
                = 
                \lambda \termType{w}{\contC{\StreamX}}.\, 
                c(x) 
                \Bigl(
                \lambda \termType{y}{X}.\, 
                u(y)
                \bigl( 
                    \lambda \termType{\pi}{\StreamX}.\, w(x\pi) \bigr)
                \Bigr).
            \end{equation*}
            This operator is a monotone map with respect to the point-wise order on $\Kleisli{\polymorphicMonotoneCont}(X,\StreamX)$ induced from the answer type lattice $\Omega$.
        \item\label{item:exe-operator-cont-monad-2} 
            The minimal and maximal $\polymorphicMonotoneCont$-execution maps exist for every $\polymorphicMonotoneCont$-coalgebra.
    \end{enumerate}
\end{myproposition}
\begin{proof}
    \refitem{item:exe-operator-cont-monad-1} follows from the formula~\refprop{prop:concrete-formula-exe-operator}.
    The monotonicity of the operator $\exeOperatorofT{\polymorphicMonotoneCont, c}$ follows from $c(x) \in \polymorphicMonotoneCont X$ for each $x \in X$.

    On \refitem{item:exe-operator-cont-monad-2}, 
    let $c$ be a $\polymorphicMonotoneCont_{\Omega}$-coalgebra.
    Since the Kleisli homset $\Kleisli{\polymorphicMonotoneCont}(X,\StreamX)$ becomes a complete lattice
    with the point-wise order induced from the order on $\Omega$, we can obtain the least and greatest fixpoints of $\exeOperatorofT{\polymorphicMonotoneCont, c}$ by the Cousot-Cousot fixpoint theorem~\cite{CousotCousot79}.
    Note that while the bottom and top elements of $\Kleisli{\monotoneContMnd}(X,\StreamX)$ are given by $\lambda\place.\, \lambda\place.\, \bot_{\Omega}$ and $\lambda\place.\, \lambda\place.\, \top_{\Omega}$, those of $\Kleisli{\affinemonotoneContMnd}(X,\StreamX)$ are given by $\lambda\place.\, \lambda w.\, \bigsqcap_{\pi \in \StreamX} w(\pi)$ and $\lambda\place.\, \lambda w.\, \bigsqcup_{\pi \in \StreamX} w(\pi)$, respectively.
\end{proof}
By~\refprop{prop:exe-operator-cont-monad},
we can always extend every continuation model for FML to one for CTL*. 
\begin{myproposition}\label{prop:min-max-CTL-continuation-model}
    Given an $\Omega$-valued continuation model $\contmodelFixptMLtuple$, we have the temporal continuation models $(c, L, \mu\exeOperatorofT{\polymorphicMonotoneCont, c})$ and $(c, L, \nu\exeOperatorofT{\polymorphicMonotoneCont, c})$,
    which we call the \emph{minimal} and \emph{maximal models} induced from $\contmodelFixptMLtuple$.
\end{myproposition}
This result is useful when applied to coalgebraic models, giving a method of extending them to temporal models. 
The following result follows from \refprop{prop:min-max-CTL-continuation-model}
and \refprop{prop:affine-trinity-lemma}.
\begin{mycorollary}\label{cor:extending-coalgebraic-model}
    Let $\TcoalgmodelFixptMLtuple$ be an $\Omega$-valued coalgebraic model where $\Diamond$ is a cartesian predicate lifting and $(\iota_X \circ c, L)$ be the induced continuation model, where $\iota$ is the monad morphism induced from $\Diamond$.
    We denote $c' := \iota_X \circ c$.
    For every $T$-execution map $u$, the minimal model and the maximal model induced from $(c', L)$ approximate the temporal coalgebraic model $\TcoalgmodelCTLtuple$:
        \begin{equation*}
            \contSemCTLsupsub{\place}{\mu\exeOperatorofT{\polymorphicMonotoneCont, c'}}{c'} 
            \sqsubseteq 
            \coalgSemCTL{\place}
            \sqsubseteq
            \contSemCTLsupsub{\place}{\nu\exeOperatorofT{\polymorphicMonotoneCont, c'}}{c'}.
        \end{equation*}
\end{mycorollary}
Note that: even if there does \emph{not exist} any $T$-execution map and thus any temporal coalgebraic model extending $\TcoalgmodelFixptMLtuple$, we always have the minimal and maximal models induced from $(c', L)$.
This gives a practical merit of transforming coalgebraic models to continuation ones in interpreting temporal logic.
\begin{myexample}[the $\Omega$-valued powerset monad and its affine part]
    For a given monad, finding its execution map is often non-trivial.
    For example, consider the $\Omega$-valued powerset monad $\OsetMnd = \Omega^{(\place)}$~\cite{Solovyov08}, whose affine part~\cite{Jacobs94} is given by $\affineOsetMnd Y = \{k \in \OsetMnd Y \mid \bigsqcup_{y \in Y} k(y) = \top\}$ for each $Y \in \sets$.  
    The monads $\OsetMnd$ and $\affineOsetMnd$ represent $\Omega$-weighted branching and its serial variant, as $\pow= \OsetMndOmega{\Bool}$ and $\nemppow = \affineOsetMndOmega{\Bool}$ represent non-determinism and serial non-determinism.
    It is not hard to see the existence of the minimal and maximal $\OsetMnd$-execution maps: we can use the Cousot-Cousot fixpoint theorem to obtain them.
    However, these maps do not necessarily restrict to $\affineOsetMnd$-execution maps for a $\affineOsetMnd$-coalgebra $c$, as the minimal $\pow$-execution map $\lambda\place.\, \bot$ does not restrict to a $\nemppow$-execution map~(\refexample{eg:pow-max-exe}).
    The existence of $\affineOsetMnd$-execution map non-trivially depends on the structures of both the lattice $\Omega$ and given coalgebras.

    Now, we apply our extension method to the monad $\affineOsetMnd$.
    First note that the monad $\OsetMnd$ is a $\monotoneContMnd$-over-monad.
    Indeed, from $\OsetMnd\terminalobj = \Omega$ by its definition, we have the Eilenberg-Moore algebra $\mu_{\terminalobj} \colon \OsetMnd \Omega = \OsetMnd\OsetMnd\terminalobj \rightarrow \OsetMnd\terminalobj = \Omega$.
    The Eilenberg-Moore algebra $\mu_{\terminalobj}$ induces a monad morphism $\iota^{\mu_{\terminalobj}} \colon \OsetMnd \Rightarrow \monotoneContMnd$ in a canonical manner~\citewithdata{Proposition~1}{Hyland+07}.
    This monad morphism also restricts to the affine part $\iota^{\mu_{\terminalobj}} \colon \affineOsetMnd \Rightarrow \monotoneContMnd$.
    Hence, by~\refcor{cor:extending-coalgebraic-model}, we obtain the under-/over-approximation of $\affineOsetMnd$-execution map in the Kleisli category $\Kleisli{\affinemonotoneContMnd}$ for every $\affineOsetMnd$-coalgebra $c$ via transferring $c$ to the $\affinemonotoneContMnd$-coalgebra $\iota^{\mu_{\terminalobj}} \circ c$.
    \refcor{cor:extending-coalgebraic-model} implies that these approximations behave as if they were the real minimal and maximal $\affineOsetMnd$-execution maps of $c$ in interpreting CTL* formulas.
\end{myexample}

\section{Weak Fixpoint Characterization under Continuation Semantics}
\label{sec:fixpt-characterization-Cont-sem-CTL}
In this section, we investigate a classical semantic property of CTL called \emph{fixpoint characterization}~\cite{EmersonHalpern85}, 
stating that CTL formulas can be encoded into FML in a way which preserves their semantics.
This property is crucial for the model-checking efficiency of CTL: 
the resulting fixpoint formulas do not contain any alternation of least and greatest fixpoint operators, and so their semantics can be computed in linear time in the formula size.
We examine the conditions for the following encoding $\epsilon$ to preserve continuation semantics.
Throughout this section, we consider only the affine monotone continuation monad $\affinemonotoneContMnd$.
\begin{mydefinition}\label{def:fixpoint-encoding}
    We define the \emph{fixpoint encoding} $\epsilon$ of the full CTL $\fullCTL$ into the full FML $\fullFixptML$ 
    by
    \begin{align*}
        \epsilon (p) 
        &:= p,
        \quad 
        &\epsilon (\lnot p) 
        &:= \lnot p,
        \\
        \epsilon (\mathsf{bb}) 
        &:= \mathsf{bb},
        \quad 
        &\epsilon (\psi_1 \star \psi_{2}) 
        &:= \epsilon\psi_1 \star \epsilon\psi_{2},
        \\
        \epsilon \bigl(\emodal\xmodal \psi\bigr) 
        &:= \Diamond (\epsilon\psi),
        \quad
        &\epsilon \bigl(\amodal\xmodal \psi\bigr) 
        &:= \Box (\epsilon\psi),
        \\
        \epsilon \Bigl(\emodal \bigl( \umodalwith{\psi_2}{\psi_1} \bigr)\Bigr) 
        &:= \mu u.\,  \epsilon\psi_1 \lor (\epsilon\psi_2 \land \Diamond u ),
        \quad
        &\epsilon \Bigl(\emodal \bigl( \wmodalwith{\psi_1}{\psi_2} \bigr)\Bigr) 
        &:= \nu u.\,  \epsilon\psi_1 \land (\epsilon\psi_2 \lor \Diamond u ),
        \\
        \epsilon \Bigl(\amodal \bigl( \umodalwith{\psi_2}{\psi_1} \bigr)\Bigr) 
        &:= \mu u.\,  \epsilon\psi_1 \lor (\epsilon\psi_2 \land \Box u ),
        \quad
        &\epsilon \Bigl(\amodal \bigl( \wmodalwith{\psi_1}{\psi_2} \bigr)\Bigr) 
        &:= \nu u.\,  \epsilon\psi_1 \land (\epsilon\psi_2 \lor \Box u )
    \end{align*}
    where $\mathsf{bb} \in \{ \ttlog,\fflog \}$ and 
    $\star \in \{ \land,\lor \}$.
    The fixpoint encoding of the $\lnot$-free CTL $\negfreeCTL$ into the $\lnot$-free fixpoint modal logic $\negfreeFixptML$ is defined as the restriction of the encoding $\epsilon$ to the subset $\negfreeCTL$.
\end{mydefinition} 
To facilitate the later statements, we introduce the following notations.
\begin{mydefinition}\label{def:Until-Wuntil-operators}
    Let $k_1, k_2 \in \contC{X}$.
    We define the continuous operators $\PhiOperator{\umodal}{k_1, k_2}, \PhiOperator{\wmodal}{k_1, k_2} \colon \contC{\StreamX} \rightarrow \contC{\StreamX}$ and
    the monotone operators
    $\PsiOperator{\umodal}{k_1, k_2}, \PsiOperator{\wmodal}{k_1, k_2} \colon \contC{X} \rightarrow \contC{X}$ by, for each $w \in \contC{\StreamX}$ and $k \in \contC{X}$,
    \begin{align*}
        \PhiOperator{\umodal}{k_1, k_2} 
        (w)
        &:= 
        \lambda \pi.\, 
        k_1 (\pi_0)
        \sqcup 
        \bigl(
            k_2 (\pi_0)
            \sqcap 
            w (\pi^{+})
        \bigr)
        &\qquad
        \PhiOperator{\wmodal}{k_1, k_2}
        (w)
        &:= 
        \lambda \pi.\, 
        k_1 (\pi_0)
        \sqcap 
        \bigl(
            k_2 (\pi_0)
            \sqcup 
            w (\pi^{+})
        \bigr)
        \\
        \PsiOperator{\umodal}{k_1, k_2} 
        (k)
        &:= 
        \lambda x.\, 
        k_1 (x)
        \sqcup 
        \bigl(
            k_2 (x)
            \sqcap 
            c(x)(k)
        \bigr)
        &\qquad
        \PsiOperator{\wmodal}{k_1, k_2}
        (k)
        &:= 
        \lambda x.\, 
        k_1 (x)
        \sqcap 
        \bigl(
            k_2 (x)
            \sqcup 
            c(x)(k)
        \bigr).
    \end{align*}
\end{mydefinition}
Under these notations, we can write down
\begin{align*}
    \contSemCTL{\emodal \umodalwith{\psi_2}{\psi_1}}
    &=
    \mu \,
    \PsiOperator{\umodal}{\contSemCTL{\psi_1}, \contSemCTL{\psi_2}}
    &\qquad
    \contSemCTL{\emodal \wmodalwith{\psi_1}{\psi_2}}
    &=
    \nu \,
    \PsiOperator{\wmodal}{\contSemCTL{\psi_1}, \contSemCTL{\psi_2}}
    \\
    \contSemFixptML{
        \mu k.\, 
        \theta_1 
        \lor 
        (
            \theta_2 
            \land 
            k
        )
        }
    &=
    \mu \,
    \PhiOperator{\umodal}{\contSemFixptML{\theta_1}, \contSemFixptML{\theta_2}}
    &\qquad
    \contSemFixptML{
        \nu k.\, 
        \theta_1 
        \land 
        (
            \theta_2 
            \lor 
            k
        )
        }
    &=
    \nu \,
    \PhiOperator{\wmodal}{\contSemFixptML{\theta_1}, \contSemFixptML{\theta_2}}
\end{align*}
for FML formulas $\theta_1, \theta_2$ and CTL formulas $\psi_1, \psi_2$.

The following additional property of $\affinemonotoneContMnd$-coalgebras and their execution maps plays a crucial role in the proof of fixpoint characterization.
\begin{mydefinition}\label{def:const-linear}
    \begin{enumerate}
        \item
            Let $Y \in \sets$. 
            A map $h \in \affinemonotoneContMnd Y$ is \emph{constant-linear} if the equations 
            \begin{equation*}
                h (\lambda \termType{y}{Y}.\, a \sqcap k(y)) 
                = a \sqcap h (k)
                \qquad
                h (\lambda \termType{y}{Y}.\, a \sqcup k(y)) 
                = a \sqcup h (k)
            \end{equation*} 
            hold for every $a \in \Omega$ and $k \in \contC{Y}$.
        \item 
            A $\affinemonotoneContMnd$-coalgebra $c \colon X \rightarrow \affinemonotoneContMnd X$ is \emph{constant-linear} if the successor $c(x) \in \affinemonotoneContMnd X$ is constant-linear for every $x \in X$.
            Similarly, we say an execution map $u \colon X \rightarrow \affinemonotoneContMnd \StreamX$ of a $\affinemonotoneContMnd$-coalgebra is \emph{constant-linear} if its values are all constant-linear.
    \end{enumerate}
\end{mydefinition}
Even if a $\affinemonotoneContMnd$-coalgebra is constant-linear, this does not necessarily imply that so are all of its execution maps.
Nonetheless, the minimal and maximal execution maps always inherit constant-linearity from coalgebras.
\begin{myproposition}\label{prop:const-linear-execution}
    Let $c \colon X \rightarrow \affinemonotoneContMnd X$ be a constant-linear $\affinemonotoneContMnd$-coalgebra. 
    The minimal and maximal execution maps $\mu \exeOperatorofT{\affinemonotoneContMnd, c}$ and $\nu \exeOperatorofT{\affinemonotoneContMnd, c}$ of $c$ are constant-linear.
\end{myproposition}
\begin{proof}
    Since the minimal and maximal execution maps $\mu\exeOperator$ and $\nu\exeOperator$ for the execution operator $\exeOperator = \exeOperatorofT{\affinemonotoneContMnd, c}$ of the monad $\affinemonotoneContMnd$ are obtained as
    \begin{equation*}
        \mu\exeOperator = \bigsqcup_{\kappa \in \Ord} 
        \exeOperator^{\kappa}(\lambda\place.\, \bot),
        \qquad
        \nu\exeOperator = \bigsqcap_{\kappa \in \Ord} 
        \exeOperator^{\kappa}(\lambda\place.\, \top)
    \end{equation*}
    by the Cousot-Cousot fixpoint theorem~\cite{CousotCousot79}, we can show the statement by the transfinite induction.
    We here see the join case for the maximal execution $\nu\exeOperator$.

    For the base step, since $\exeOperator^{0}(\lambda\place.\, \top)=\lambda x.\, \lambda w.\, \bigsqcup_{\pi \in \StreamX} w(\pi)$, we have
    \begin{equation*}
        \exeOperator^{0}(\lambda\place.\, \top)(x)(a \sqcap w)
        =
        \bigsqcup_{\pi \in \StreamX} a \sqcap w(\pi)
        =
        a \sqcap \bigsqcup_{\pi \in \StreamX} w(\pi)
        =
        a \sqcap \exeOperator^{0}(\lambda\place.\, \top)(x)(w)
    \end{equation*}
    for each $x \in X$, $w \in \contC{\StreamX}$ and $a \in \Omega$.
    For each successor ordinal $\kappa + 1$, assume 
    $\exeOperator^{\kappa}(\lambda\place.\, \top)(x)(a \sqcap w) = a \sqcap \exeOperator^{\kappa}(\lambda\place.\, \top)(x)(w)$
    for every $x \in X$, $w \in \contC{\StreamX}$ and $a \in \Omega$.
    By~\refitem{item:exe-operator-cont-monad-1} of~\refprop{prop:exe-operator-cont-monad}, we have 
    \begin{align*}
        \exeOperator^{\kappa +1}(\lambda\place.\, \top)(x)(a \sqcap w)
        &=
        \exeOperator(\exeOperator^{\kappa})(\lambda\place.\, \top)(x)(a \sqcap w)
        \\
        &=
        c(x)
        \Bigl(
            \lambda y.\,
            \exeOperator^{\kappa}(\lambda\place.\, \top)(y)
            \bigl(
                \lambda \pi.\,
                a \sqcap w(x\pi)
            \bigr)
        \Bigr)
        \\
        &=
        c(x)
        \Bigl(
            \lambda y.\,
            \exeOperator^{\kappa}(\lambda\place.\, \top)(y)
            \bigl(
                \lambda \pi.\,
                a \sqcap w_x(\pi)
            \bigr)
        \Bigr)
    \end{align*}
    where we denote $w_x = \lambda \pi.\, w(x\pi)$. 
    By applying the induction hypothesis to this $w_x \in \contC{\StreamX}$, we obtain 
    \begin{align*}
        \exeOperator^{\kappa +1}(\lambda\place.\, \top)(x)(a \sqcap w)
        &=
        c(x)
        \Bigl(
            \lambda y.\,
            \exeOperator^{\kappa}(\lambda\place.\, \top)(y)
            \bigl(
                \lambda \pi.\,
                a \sqcap w_x(\pi)
            \bigr)
        \Bigr)
        \\
        &=
        c(x)
        \Bigl(
            \lambda y.\,
            a \sqcap 
            \exeOperator^{\kappa}(\lambda\place.\, \top)(y)
            \bigl(
                \lambda \pi.\,
                w_x(\pi)
            \bigr)
        \Bigr)
        \\
        &=
        a \sqcap 
        c(x)
        \Bigl(
            \lambda y.\,
            \exeOperator^{\kappa}(\lambda\place.\, \top)(y)
            \bigl(
                \lambda \pi.\,
                w_x(\pi)
            \bigr)
        \Bigr)
        \\
        &=
        a \sqcap
        \exeOperator^{\kappa +1}(\lambda\place.\, \top)(x)(w).
    \end{align*}
    Here, we used $c(x)$ being constant-linear.
    Finally, for each limit ordinal $\lambda$, assume $\exeOperator^{\kappa}(\lambda\place.\, \top)$ is constant-linear for every ordinal $\kappa<\lambda$.
    Under this induction hypothesis, we conclude
    \begin{equation*}
        \exeOperator^{\lambda}(\lambda\place.\, \top)(x)(a \sqcap w)
        =
        \bigsqcap_{\kappa < \lambda} 
        \exeOperator^{\kappa}(\lambda\place.\, \top)(x)(a \sqcap w)
        =
        \bigsqcap_{\kappa < \lambda} 
        a \sqcap 
        \exeOperator^{\kappa}(\lambda\place.\, \top)(x)(w)
        =
        a \sqcap 
        \bigsqcap_{\kappa < \lambda} 
        \exeOperator^{\kappa}(\lambda\place.\, \top)(x)(w)
    \end{equation*}
    for every $x \in X$, $w \in \contC{\StreamX}$ and $a \in \Omega$.
    Since $\bigsqcap_{\kappa < \lambda} \exeOperator^{\kappa}(\lambda\place.\, \top) = \exeOperator^{\lambda}(\lambda\place.\, \top)$, we finally obtain $\exeOperator^{\lambda}(\lambda\place.\, \top)(x)(a \sqcap w) = a \sqcap \exeOperator^{\lambda}(\lambda\place.\, \top)(x)(w)$.
\end{proof}
The fixpoint characterization result~(\refitem{item:original-fixpoint-char} below) and its weaker version~(\refitem{item:weak-fixpoint-char} below) of CTL under continuation semantics hold under the constant-linearity assumption.
\begin{mytheorem}[(weak) fixpoint characterization for CTL]\label{thm:fixpoint-characterization}
    Let $\contmodelCTLtuple$ be an $\Omega$-valued temporal continuation model with constant-linear maps $c$ and $u$.
    Let $\psi$ be a CTL formula.
    \begin{enumerate}
        \item\label{item:weak-fixpoint-char} 
            When the execution map $u$ is constant-linear, the following equality and inequalities hold for the fixpoint encoding $\epsilon$~(\refdef{def:fixpoint-encoding}).
            \begin{enumerate}
                \item\label{item:weak-fixpoint-char-1} 
                If the formula $\psi$ does not contain the symbols $\umodal$ nor $\wmodal$, the equality 
                $\contSemCTL{\psi} = \contSemFixptML{\epsilon\psi}$ holds.
                \item\label{item:weak-fixpoint-char-2} 
                If the formula $\psi$ contains only the symbol $\umodal$, the inequality 
                $\contSemCTL{\psi} \sqsupseteq \contSemFixptML{\epsilon\psi}$ holds.
                \item\label{item:weak-fixpoint-char-3} 
                If the formula $\psi$ contains only the symbol $\wmodal$, the inequality 
                $\contSemCTL{\psi} \sqsubseteq \contSemFixptML{\epsilon\psi}$ holds.
            \end{enumerate}
        \item\label{item:original-fixpoint-char} 
            Moreover, the inequalities above become equalities if the inequalities 
            \begin{equation*}
                \lambda x.\,
                u(x)
                \bigl(
                    \mu \,
                \PhiOperator{\umodal}{\contSemCTL{\psi_1}, \contSemCTL{\psi_2}}
                \bigr)
                \sqsubseteq
                \mu \,
                \PsiOperator{\umodal}{\contSemFixptML{\epsilon\psi_1}, \contSemFixptML{\epsilon\psi_1}}
                \qquad
                \lambda x.\,
                u(x)
                \bigl(
                    \nu \,
                    \PhiOperator{\wmodal}{\contSemCTL{\psi_1}, \contSemCTL{\psi_2}}
                \bigr)
                \sqsupseteq
                \nu \,
                \PsiOperator{\wmodal}{\contSemFixptML{\epsilon\psi_1}, \contSemFixptML{\epsilon\psi_2}}
            \end{equation*}
            hold for $\psi$-sub-formulas $\psi_1, \psi_2$.
            Here, the operators $\PsiU,\PsiW$ and $\PhiU,\PhiW$ are as in~\refdef{def:Until-Wuntil-operators}.
    \end{enumerate}
\end{mytheorem}
\begin{myremark}[applicablility of weak fixpoint characterization]
    We expect our weaker version of the fixpoint characterization~(\refitem{item:weak-fixpoint-char} of \refthm{thm:fixpoint-characterization}) still has several practical values.
    When one uses the logic CTL as a specification-description language, the verifier is often concerned with not only the precise evaluation but also with ``under-approximations'' of one's desired specification.
    Our weaker result~(\refitem{item:weak-fixpoint-char-2}) implies liveness properties can be approximated by fast fixpoint-based algorithms, which can terminate in linear time in the number of sub-formulas of the specification in question.
    
    We can also use the condition in~\refitem{item:original-fixpoint-char} as criteria to make an efficient choice of execution map with respect to one's objective system.
    When one luckily finds a join/meet-continuous execution map, our result guarantees verification of liveness/safety properties to be calculated in linear-time.
    Even when it is difficult to find such a well-behaving map in lattice theoretic terms, there is still room for ad-hoc approaches for specific systems or arranging the system itself to fit with the specification.
\end{myremark}
As a special case of~\refthm{thm:fixpoint-characterization}, the minimal and maximal models~(\refprop{prop:min-max-CTL-continuation-model}) with constant-linear $c$ always enjoy the weak fixpoint characterization by~\refprop{prop:const-linear-execution}.
\begin{myproposition}\label{prop:fixpt-char-instances}
    For every $\Omega$-valued continuation model $\contmodelFixptMLtuple$ with constant-linear $c$, the minimal and maximal models $(c, L, \mu \exeOperatorofT{\affinemonotoneContMnd, c})$ and $(c, L, \nu \exeOperatorofT{\affinemonotoneContMnd, c})$
    enjoy the weak fixpoint characterization.
\end{myproposition}
\begin{myexample}
    When the lattice $\Omega$ is the booleans $\Bool$, every $\affinemonotoneContMnd_{\Bool}$-coalgebra for the affine monotone continuation monad $\affinemonotoneContMnd_{\Bool}$ can be easily checked to be constant-linear.
    Thus, \refprop{prop:fixpt-char-instances} implies that the minimal and maximal ones of $\Bool$-valued temporal continuation model
    enjoy the weak fixpoint characterization.
    Moreover, we can show that the maximal model also satisfies the inequalities in~\refitem{item:original-fixpoint-char} of~\refthm{thm:fixpoint-characterization} and thus enjoys the (full) fixpoint characterization.
    The proof for this fixpoint characterization result goes essentially the same as that of \emph{Alternating-time Temporal Logic}~\cite{BaierKatoen08}.
\end{myexample}

We show~\refthm{thm:fixpoint-characterization}.
The core parts of the proof are the following facts.
\begin{mylemma}\label{lem:facts-for-weak-fixpt-ch}
    Let $c \colon X \rightarrow \affinemonotoneContMnd X$ be a $\affinemonotoneContMnd$-coalgebra, $u \colon X \rightarrow \affinemonotoneContMnd \StreamX$ be an execution map of $c$.
    \begin{enumerate}
        \item\label{item:affine-trace-equations-Cont} 
        For every $k \in \contC{X}$ and $x \in X$, we have
        \begin{equation*}
            u(x)
            (
                \lambda \pi.\,
                k(\pi_0)
            )
            =
            k(x) 
            \qquad
            \text{ and }
            \qquad
            u(x)
            (
                \lambda \pi.\,
                k(\pi_1)
            )
            =
            c(x)(k).
        \end{equation*} 
        \item\label{item:Pre-Post-fixpt-Psi}  
        Assume the maps $c$ and $u$ are moreover constant-linear.
        Then, for every $k_1,k_2 \in \contC{X}$, $w \in \contC{\StreamX}$ and $x \in X$, we have 
        the equations
        \begin{equation*}
            \PsiOperatorU{k_1,k_2}
            \bigl(
                \lambda y.\, u(y)(w)
            \bigr)
            =
            u(x)
            \bigl(
                \PhiOperatorU{k_1,k_2}
                (w)
            \bigr),
            \qquad
            \PsiOperatorW{k_1,k_2}
            \bigl(
                \lambda y.\, u(y)(w)
            \bigr)
            =
            u(x)
            \bigl(
                \PhiOperatorW{k_1,k_2}
                (w)
            \bigr)
        \end{equation*}
        where the operators $\PsiU,\PsiW$ and $\PhiU,\PhiW$ are the ones defined in~\refdef{def:Until-Wuntil-operators}.
    \end{enumerate}
\end{mylemma}
\begin{proof}
    We denote $\exeOperator := \exeOperatorofT{\affinemonotoneContMnd,c}$.
    Since $u$ is an execution map of $c$, the equality $u = \exeOperator(u)$ holds.
    Thus, the first equation of \refitem{item:affine-trace-equations-Cont} follows from the following calculation:
    \begin{align*} 
        u(x)
        \bigl( 
            \lambda \pi.\,
            k (\pi_0)
        \bigr)
        &=
        \exeOperator(u)(x)
        \bigl( 
            \lambda \pi.\,
            k (\pi_0)
        \bigr)
        \\
        &=
        c(x) 
        \bigl(
                \lambda y.\, 
                u(y)
                \bigl( 
                    \lambda \pi'.\,
                    (\lambda \pi.\,
                    k (\pi_0))
                    (x\pi') 
                \bigr)
        \bigr)
        \\
        &=
        c(x) 
        \bigl(
                \lambda y.\, 
                u(y)
                \bigl( 
                    \lambda \pi'.\,
                    k (x)
                \bigr)
        \bigr)
        \\
        &=
        k (x)
    \end{align*}
    where the last transformation comes from the affineness of $c$ and $u$.
    The second equation of~\refitem{item:affine-trace-equations-Cont} comes from the equality $u = \exeOperator(u)$ and also the first equation of~\refitem{item:affine-trace-equations-Cont}:
    \begin{align*} 
        u(x)
        \bigl( 
            \lambda \pi.\,
            k (\pi_1)
        \bigr)
        &= 
        \exeOperator(u)(x) 
        \bigl( 
            \lambda \pi.\,
            k (\pi_1)
        \bigr)
        \\
        &=
        c(x) 
        \Bigl(
                \lambda y.\, 
                u(y)
                \bigl( 
                    \lambda \pi'.\,
                    (\lambda \pi.\,
                    k (\pi_1))
                    (x\pi) 
                \bigr)
        \Bigr)
        \\
        &=
        c(x) 
        \bigl(
                \lambda y.\, 
                u(y)
                \bigl( 
                    \lambda \pi'.\,
                    k (\pi'_0)
                \bigr)
        \bigr)
        \\
        &=
        c(x) 
        \bigl(
                \lambda y.\, 
                k (y)
        \bigr)
        \\
        &=
        c(x) 
        (k).
    \end{align*}

    We next show \refitem{item:Pre-Post-fixpt-Psi} only for the $\umodal$ case, since the $\wmodal$ case is its dual.
    We just denote $\Psi := \PsiOperatorU{k_1,k_2}$ and $\Phi := \PhiOperatorU{k_1,k_2}$ here.
    For each $x \in X$ and $w \in \contC{\StreamX}$, we have
    \begin{align*}
        u(x)
        \bigl(
            \Phi(w)
        \bigr) 
        &= 
        \exeOperator(u)(x)
        \bigl(
            \Phi(w)
        \bigr)
        \\
        &=
        c(x) 
        \bigl(
            \lambda y.\, 
                u(y)
                \bigl( 
                    \lambda \pi.\, \Phi(w)(x\pi) 
                \bigr)
        \bigr)
        \\
        &=
        c(x) 
        \Bigl(
            \lambda y.\, 
                u(y)
                \bigl( 
                    \lambda \pi.\, 
                    k_1(x) 
                    \sqcup 
                    \bigl(
                        k_2(x) \sqcap (w)(\pi)
                    \bigr) 
                \bigr)
        \Bigr)
        \\
        &=
        k_1(x) 
        \sqcup
        \Bigl(
            k_2(x) 
            \sqcap
            c(x)
            \bigl(
                \lambda y.\, 
                u(y)
                (  
                    w
                )
            \bigr)
        \Bigr)
        \\
        &=
        \Psi
        \bigl(
            \lambda y.\, 
            u(y)
            (  
                w
            )
        \bigr)
        (x)
    \end{align*}
    where we used constant-linearity of $c, u$.
    Thus, we conclude $\lambda x.\, u(x)\bigl(\Phi (w)\bigl) = \Psi
    \bigl(
        \lambda x.\, 
        u(x)
        (  
            w
        )
    \bigr)$.
\end{proof}
\begin{proof}[Proof of \refthm{thm:fixpoint-characterization}]
    We first show \refitem{item:weak-fixpoint-char-1}: the equality $\contSemCTL{\psi} = \contSemFixptML{\epsilon\psi}$
    for the formula $\psi$ which does not contain the symbols $\umodal$ nor $\wmodal$.
    The proof goes by induction.
    When the formula $\psi$ is an atomic predicate or boolean expression of sub-formulas, the above equality immediately follows from the definition of the fixpoint encoding $\epsilon$~(\refdef{def:fixpoint-encoding}) and the first equation of \refitem{item:affine-trace-equations-Cont} of~\reflem{lem:facts-for-weak-fixpt-ch}.

    For the case $\psi = \emodal\xmodal \psi'$, the induction hypothesis asserts $\contSemCTL{\psi'} = \contSemFixptML{\epsilon\psi'}$.
    Let $x \in X$.
    By the definition of $\contSemCTL{\place}$~(\refdef{def:continuation-semantics-CTL}), we have
    $\contSemCTL{\emodal\xmodal \psi'}(x)
    = u(x) (\contSemCTL{\xmodal \psi'})
    = u(x) 
    \bigl( 
        \pull{\zeta_2} \circ \pull{\zeta_1}(\contSemCTL{\psi'})
    \bigr)$.
    Since we have
    $\pull{\zeta_2} \circ \pull{\zeta_1}(\contSemCTL{\psi'})
    =
    \lambda \pi.\,
    \contSemCTL{\psi'}
    \bigl(
        \zeta_1 \circ \zeta_2 (\pi)
    \bigr)
    =
    \lambda \pi.\,
    \contSemCTL{\psi'} (\pi_1)$ by the definition of the final $\PathFunctorX$-coalgebra map $\zeta$,
    we obtain
    $\contSemCTL{\psi}(x)
    = u(x) 
    \bigl( 
        \lambda \pi.\,
        \contSemCTL{\psi'} (\pi_1)
    \bigr)$.
    Finally, by the induction hypothesis $\contSemCTL{\psi'} = \contSemFixptML{\epsilon\psi'} \in \contC{X}$ and the second equation of~\refitem{item:affine-trace-equations-Cont} of~\reflem{lem:facts-for-weak-fixpt-ch} of~\reflem{lem:facts-for-weak-fixpt-ch}, we conclude
    \begin{equation*}
        \contSemCTL{\emodal\xmodal \psi'}(x)
        = u(x) 
        \bigl( 
            \lambda \pi.\,
            \contSemCTL{\psi'} (\pi_1)
        \bigr)
        = u(x) 
        \bigl( 
            \lambda \pi.\,
            \contSemFixptML{\epsilon\psi'} (\pi_1)
        \bigr)
        = 
        c(x)
        (
            \contSemFixptML{\epsilon\psi'} 
        )
        =
        \contSemFixptML{\Diamond(\epsilon\psi')}(x) 
        =
        \contSemFixptML{\epsilon(\emodal\xmodal\psi')}(x).
    \end{equation*}

    Next, we prove \refitem{item:weak-fixpoint-char-2} for the $\umodal$ case by induction: the proof for the $\wmodal$ case is obtained dually.
    Since we have seen the inequality (indeed equality) $\contSemCTL{\psi} \sqsupseteq \contSemFixptML{\epsilon\psi}$ for each CTL formula $\psi$ without symbol $\umodal$, it suffices to show the inequality $\contSemCTL{\psi} \sqsupseteq \contSemFixptML{\epsilon\psi}$ for the CTL formula $\psi = \emodal(\umodalwith{\psi_2}{\psi_1})$ under the induction hypotheses $\contSemCTL{\psi_1} \sqsupseteq \contSemFixptML{\epsilon\psi_1}$ and $\contSemCTL{\psi_2} \sqsupseteq \contSemFixptML{\epsilon\psi_2}$.
    Let us denote $k_i := \contSemCTL{\psi_i}$ and $k'_i := \contSemFixptML{\epsilon\psi_i}$ for $i = 1,2$.
    Using the operators defined in~\refdef{def:Until-Wuntil-operators}, the inequality in question amounts to  
    \begin{equation}
        \label{ineq:pre-fixpt-Psi-U}
        \lambda x.\,
                u(x)
                \bigl(
                    \mu \,
                \PhiOperator{\umodal}{k_1, k_2}
                \bigr)
                \sqsupseteq
                \mu \,
                \PsiOperator{\umodal}{k'_1, k'_2}
    \end{equation}
    for $k_i,k'_i$ with $k_i \sqsupseteq k'_i$ for $i = 1,2$.

    Since the operator $\PsiOperator{\umodal}{k'_1, k'_2}$ is a monotone map over the complete lattice $\contC{X}$, we can use the Knaster-Tarski fixpoint theorem~\cite{Tarski55}, which implies $\mu \, \PsiOperator{\umodal}{k'_1, k'_2}$ is the least one among all the pre-fixpoints of the operator $\PsiOperator{\umodal}{k'_1, k'_2}$.
    Thus, to see inequality~\ref{ineq:pre-fixpt-Psi-U}, it suffices to show that $\LHS$ of inequality~\ref{ineq:pre-fixpt-Psi-U},
    $\lambda x.\, 
    u(x) 
    \bigl(
        \mu \,
        \PhiOperator{\umodal}{k_1, k_2}
    \bigr)$,
    is a pre-fixpoint of $\PsiOperator{\umodal}{k'_1, k'_2}$.
    This can be seen the following calculation, using \refitem{item:Pre-Post-fixpt-Psi} of~\reflem{lem:facts-for-weak-fixpt-ch},
    \begin{equation*}
        \PsiOperator{\umodal}{k'_1, k'_2}
        \Bigl(
            \lambda x.\, 
            u(x) 
            \bigl(
                \mu \,
                \PhiOperator{\umodal}{k_1, k_2}
            \bigr)
        \Bigr)
        =
        \lambda x.\, 
        u(x)
        \Bigl(
            \PhiOperator{\umodal}{k'_1, k'_2} 
            \bigl(
                \mu \,
                \PhiOperator{\umodal}{k_1, k_2}
            \bigr)
        \Bigr)
        \sqsubseteq
        \lambda x.\, 
        u(x)
        \Bigl(
            \PhiOperator{\umodal}{k_1, k_2} 
            \bigl(
                \mu \,
                \PhiOperator{\umodal}{k_1, k_2}
            \bigr)
        \Bigr)
        =
        \lambda x.\, 
        u(x) 
        \bigl(
            \mu \,
            \PhiOperator{\umodal}{k_1, k_2}
        \bigr).
    \end{equation*}
    Here note that $\PhiOperator{\umodal}{k_1, k_2}$ is monotone with respect to $k_1, k_2$.
    Thus, we conclude $\LHS$ of inequality~\ref{ineq:pre-fixpt-Psi-U} is a pre-fixpoint of $\PsiOperator{\umodal}{k'_1, k'_2}$, and thus we have $\LHS \sqsubseteq \mu \PsiOperator{\umodal}{k'_1, k'_2} = \RHS$.

    \refitem{item:original-fixpoint-char} immediately follows from~\refitem{item:weak-fixpoint-char}: the inequalities in~\refitem{item:original-fixpoint-char}, combined with inequality~\ref{ineq:pre-fixpt-Psi-U} (and its dual) imply $\contSemCTL{\psi} = \contSemFixptML{\epsilon\psi}$ for the CTL formulas $\psi = \emodal(\umodalwith{\psi_2}{\psi_1})$ and $\psi = \emodal(\wmodalwith{\psi_1}{\psi_2})$ under the induction hypotheses $\contSemCTL{\psi_1} = \contSemFixptML{\epsilon\psi_1}$ and $\contSemCTL{\psi_2} = \contSemFixptML{\epsilon\psi_2}$.
\end{proof}

\section{Conclusion}
We have shown how continuation semantics can be used to unify modal and temporal
coalgebraic logics whose domain of truth values is a complete lattice. One advantage
of this unified perspective is that, in contrast to coalgebraic models, modal
continuation models can always be extended to temporal ones using either maximal or
minimal execution maps. (The difficulty in finding maximal trace/execution maps for
monads seems to be the main reason why few concrete examples of execution maps are
known.) Additionally, while the category-theoretic perspective to modelling reactive
systems has the benefit of uniformity, a lattice-theoretic perspective has the
potential to ease the development and analysis of verification algorithms, as
witnessed by work in \cite{Hasuo+16,Baldan+19}.
Our continuation-based approach can utilize both of these two perspectives.


Our continuation semantics casts new light on \emph{non-commutative} aspect of formal verification, which seems to be less investigated in both the classical and meta-theoretic/coalgebraic frameworks of verification.
Indeed, practically important properties like model-checking efficiency are not restricted to commutative, or concurrent, settings like non-deterministic and probabilistic ones, as we observed in our (weak) fixpoint characterization result under continuation semantics.
Existing coalgebraic formulations of formal verification potentially encompass non-commutative settings by considering non-commutative monads like continuation monads as their instances, but it is rather rare that such non-commutative monads are treated as main subjects, or even as examples, of the formulations.
We will further clarify to what extent commutativity should be assumed in various results and constructions of verification techniques, and what methods can be applied to non-commutative systems modelled as coalgebras of continuation monads.

As a concrete direction of our future work, we will elaborate on concrete sub/over-monads of continuation monads for complete
lattices other than the boolean one, like the expectation monad~\cite{JacobsMandemakerFurber16} and the $\Omega$-powerset monads, and execution maps that are neither minimal nor
maximal. For the latter, we believe a finer hierarchy of execution maps could be
used to account for additional types of temporal behavior, beyond the extreme cases
of finite and infinite traces.
For example, we expect this hierarchy would capture coalgebraically formulated memory-full/memory-less strategies of alternating 2-player games~\cite{PlummerCirstea25}.

Another future direction is an extension of continuation semantics to dynamic logics like PDL~\cite{FischerLadner79} and Game Logic~\cite{Parikh85}.
This extension comes naturally when we consider many-sorted systems, or ``\emph{computations}'', instead of (single-sorted) systems.
We expect our continuation semantics can be defined in the same manner also over these many-sorted models, and would yield semantic equivalence to coalgebra-based semantics for dynamic logics~\cite{HansenKupkeLeal14,HansenKupke15}.
By this extension, we hope to provide a unified framework for both system verification based on temporal logics and program verification based on dynamic logics, such as categorical Hoare logic~\cite{AguirreKatsumata20}.

\section*{Acknowledgment}
Thanks to Masahito Hasegawa for discussion on the notion of continuation and related work and the reviewers for helpful comments and suggestions to improve our paper. 
Corina Cirstea was funded by a Leverhulme Trust Research Project Grant (RPG-2020-232).



\bibliographystyle{./entics}
\bibliography{main}

\end{document}